\documentclass[structabstract]{aa}  
\usepackage{graphicx}
\usepackage{txfonts}
\begin{document}
   \title{The evolution of cool-core clusters}


   \author{J.S. Santos,
          \inst{1}
          P. Tozzi,\inst{1}
          P. Rosati,\inst{2}
	  H. B\"ohringer\inst{3}
	}

   \institute{
	     \inst{1} INAF, Osservatorio Astronomico di Trieste, via G.B. Tiepolo 11, 
34131, Trieste, Italy \\
             \email{jsantos@oats.inaf.it} \\
             \inst{2} European Southern Observatory, Karl-Schwarzchild Strasse 2, 85748 
Garching, Germany \\
	     \inst{3} Max-Planck-Institut f\"ur extraterrestrische Physik,
              Giessenbachstra\ss e, 85748 Garching, Germany\\
             }

   \date{Received ... ; accepted ...}


  \abstract
   {Cool-core clusters are characterized by strong surface brightness peaks in the X-ray 
    emission from the Intra Cluster Medium (ICM). This phenomenon is associated with 
    complex physics in the ICM and has been a subject of intense debate and 
    investigation in recent years. 
    The evolution of cool-cores is still poorly constrained because of
   the small sample statistics and the observational challenge of analysing
    high redshift clusters.}
  {In order to quantify the evolution in the cool-core cluster population, 
we robustly measure the cool-core strength in a local, representative cluster sample, 
and in the largest sample of high-redshift clusters available to date.
   }
  {We use high-resolution \textit{Chandra} data of three representative 
cluster samples spanning different redshift ranges: (i) the low redshift 
sample from the 400 Square degree (SD) survey with median $\langle z \rangle = 0.08$, (ii)
the high redshift sample from the 400 SD Survey with median $\langle z
\rangle =0.59$, and (iii) 15 clusters drawn from the Rosat Distant
Cluster Survey and the Wide Angle Rosat Pointed Survey, with median
$\langle z \rangle = 0.83$.  Our analysis is based on the measurement of
the surface brightness concentration, $c_{SB}$ (Santos et al 2008),
which allows us to characterize the cool-core strength in low signal-to-noise data. 
We also obtain gas density profiles to derive cluster central cooling times and entropy.  
In addition to the X-ray analysis, we search for radio counterparts associated with 
the cluster cores. }
   {We find a statistically significant difference in the $c_{SB}$ distributions of 
the two high-$z$ samples, pointing towards a lack of concentrated clusters 
in the 400 SD high-$z$ sample. 
Taking this into account, we confirm a negative evolution in the fraction of 
cool-core clusters with redshift, in particular for very strong cool-cores. 
This result is validated by the central entropy and central cooling time, which show 
strong anti-correlations with $c_{SB}$.
However, the amount of evolution is significantly smaller than previously claimed, 
leaving room for a large population of well formed cool-cores at $z\sim1$.
Finally, we explore the potential of the proposed X-ray mission Wide Field X-ray Telescope 
to detect and quantify cool-cores up to $z$=1.5.}
   {}

   \authorrunning{J.S.Santos et al.}
   \titlerunning{The evolution of cool-core clusters}

   \maketitle

%

\section{Introduction}

X-ray observations show that the majority of local clusters have a prominent 
central surface brightness peak. In addition, the inferred cooling time of the 
intracluster medium in the core region is much shorter than the 
dynamical time of the cluster, implying the presence of a cooling flow
(Fabian et al. \cite{fabian94}). However, the minimum temperature in the center is only 
a factor of $\sim 3$ lower than the ambient temperature, therefore the gas does not 
appear to cool massively to low temperatures.
The properties and the formation mechanism of these cool-cores 
(CC) are an open problem which forces one to consider complex non-gravitational 
physical processes in order to provide smoothly distributed heating on scales 
of about 100 kpc.  A successful model is expected to include phenomena such as 
radiative cooling, heating by a central radio source, thermal conduction or other 
forms of feedback (see Peterson \& Fabian \cite{peterson05} and references therein).

The impact of cool-cores on the local cluster population has been
extensively studied for over a decade (Peres et al. \cite{peres}). Occasionally, 
the cool-core phenomenon is described as a bimodal feature, with a clear 
separation between cool-cores and non cool-core clusters on the basis of 
the central value of the ICM entropy (e.g. Sanderson et al. \cite{sanderson}, 
Cavagnolo et al. \cite{cav}). Concurrently though, several works have 
distinguished between three regimes 
of cooling, with an intermediate category bridging the extremes (Leccardi et al. \cite{leccardi}, 
Bauer et al. \cite{bauer}, Morandi \& Ettori \cite{morandi}). 
Contrary to the bimodal scenario, 
this approach suggests a gradual transition from the non cool-core to the
 cool-core status, with profound implications on the time scale and thus the 
nature of the heating mechanism.

A crucial aspect here is a proper definition of cool-core, since a variety of 
methods or cooling estimators have been proposed to classify and quantify cool 
cores in local clusters, producing results that occasionally lead to different 
interpretations. A comprehensive review and comparison of 16 different cooling estimators
has been recently published by Hudson et al. (\cite{hudson}), using the local sample HIFLUCGS 
(Reiprich \& B\"ohringer \cite{reip}). In this work, the central cooling time was selected 
as the most efficient probe to quantify cool-cores.

The quality of the cooling estimator is an important issue to consider 
because it depends significantly on the signal-to-noise ratio of the X-ray data, 
and therefore changes considerably for high-redshift clusters with respect to 
local ones. This aspect is particularly relevant since the evolution of cool-core clusters 
is a major piece of information in order to constrain the cool-core physics.

X-ray observations have established that cool-cores dominate the local
cluster population, with an abundance of 50 to 70\% (e.g. Chen et al. \cite{chen}, 
Dunn \& Fabian \cite{dunn}, Hudson et al. \cite{hudson}).  
The evolution of cool-cores has been measured up to redshift 0.4. Using the high-$z$ end 
of the BCS sample Bauer et al. (\cite{bauer}) concluded that the fraction of cool-cores does not 
significantly evolve up to $z\thicksim~0.4$. Their results, based on spatially resolved 
spectroscopy, showed that clusters in this redshift range have the same temperature decrement 
(about one-third), as the nearby CC's, and in addition, their central cooling times are similar. 

 At present, about twenty X-ray clusters with $z>1$ have been confirmed. While 
most of them were detected in ROSAT surveys, a significant fraction of serendipitous 
high-$z$ clusters has been added in recent years thanks to the XMM-Newton archive.  
Nevertheless, the largest distant cluster samples with sufficiently deep follow-up observations 
are still the ones from ROSAT. Therefore, the study of the presence 
of cool-cores at redshift greater than 0.5 is plagued by low statistics, and currently is 
limited to two works. Using the 400 Square Degree Survey (hereafter 400 SD, Burenin et al. \cite{burenin}) 
which reaches $z=0.9$, Vikhlinin et al. (\cite{vikhlinin06}) concluded, on the basis of 
a cuspiness parameter defined as the logarithmic derivative of the density profile, 
that there is a lack of cool-core clusters at 0.5$<z<$0.9, with respect to the local cluster population. 
They propose that such a strong negative evolution may be related to the higher 
cluster merging rate expected at these redshifts.

In Santos et al. (\cite{joana}), we 
adopted a simple diagnostic based on the concentration of the surface brightness 
(which strongly anti-correlates with the central cooling time), and 
measured the fraction of cool-cores out to the current redshift limit ($z \sim 1.4$) using 
mostly the RDCS sample.  At variance with previous results, we found a significant 
fraction of what we term moderate cool-cores. To clarify these results we need a more 
detailed investigation as we propose in this Paper.

We present a comprehensive analysis of the characterization and 
abundance of cool-cores across the entire cluster population, out to the 
current highest redshift where clusters have been detected in the
 X-ray band. To this aim we analysed 3 representative samples corresponding to 
three different redshift ranges, using only \textit{Chandra} data. 
The high-resolution and low background of \textit{Chandra} are the key features that 
distinguish it as the only X-ray observatory capable of unveiling the small cores of 
distant clusters. Our analysis, based on the 
observed surface brightness and a simple concentration parameter (Santos et al. \cite{joana}), 
is uniform and robust over the entire explored range of redshifts.  This investigation 
enables us to assess the impact of detection bias in the high-$z$ cluster samples, and to 
measure on solid ground the evolution of cool-core cluster distribution.
Furthermore, we investigate the presence of radio sources associated with the cluster 
cores by exploring the NVSS archive.  
Finally, we assess the potential of the next generation X-ray survey mission Wide 
Field X-ray Telescope (WFXT, Giacconi et al. \cite{giacconi}) in measuring the cool-core evolution.

The paper is organized as follows: in \S 2 and \S 3 we describe
  the low and high redshift samples, respectively.  In \S 4 we
  describe the data reduction and analysis, devoting \S 5 to a careful
  investigation of the surface brightness concentration parameter. 
In \S 6 we perform surface brightness fits to the data using the 
well-known $\beta$-models. The entropy profiles and a histogram of K20 are 
discussed in \S 7, and the cooling time distribution follows in \S 8. 
In \S 9 we investigate the presence of radio sources in cool-core and non 
cool-core clusters. \S 10 is devoted to the future perspective with a 
next-generation X-ray survey mission. Our conclusions are summarized in \S 11.  
The cosmological parameters used throughout the paper are: $H_{0}$=70 km/s/Mpc,
$\Omega_{\Lambda}$=0.7 and $\Omega_{\rm m}$=0.3.

\section{The low redshift sample}

The measured fraction of local cool-core clusters 
ranges from 70-90\% (Peres et al. \cite{peres}, from the B55 sample),
to 49\% (Chen et al. \cite{chen}, from the HIFLUGCS sample), 
45\% (Sanderson et al. \cite{sanderson}, from a subsample of HIFLUGCS), and more recently
44\%-72\% (Hudson et al. \cite{hudson}, from the HIFLUGCS sample).
The differences in these fractions are  mostly due to different definitions of cool-core.  
  To date, Hudson et al. (\cite{hudson}) is the only exhaustive work that explores the 
  definition of cool-core and compares the currently used methods to quantify it. After 
  examining 16 different cool-core diagnostics, they 
  conclude  that the central cooling time is the best parameter for low redshift clusters with high 
  quality data, and the surface brightness concentration, $c_{SB}$, presented in 
  Santos et al. (\cite{joana}), is chosen as the best parameter for distant clusters.
  These results reinforce our strategy to apply the same concentration diagnostics
  to local and distant cluster samples, and compare directly the
  measured distributions, avoiding the use of a sharp threshold in the
  definition of the cool-core strength.

The local cluster sample used in this work is drawn form the
catalog of the 400 Square Degree (SD) Survey (Burenin et al. \cite{burenin}), an 
X-ray survey which detected 266 confirmed galaxy clusters, groups or 
individual elliptical galaxies out to $z \sim 1$ using archival ROSAT 
PSPC observations. The sample is complete down to a flux limit of 
$1.4 \times 10^{-13}$ erg s$^{-1}$ cm$^{-2}$.
We extract a subsample of clusters observed with {\sl Chandra} with $z>$0.05, 
in order to be able to sample the surface brightness profiles out to a radius of 
400 kpc within the field of view. Hence, our local sample spans the redshift
range [0.05 - 0.217] (with a median redshift $\langle z \rangle = 0.08$) and 
includes 28 clusters (see Table 1 and Fig. 1). The well-known clusters Hydra-A 
and S1101, which show a very strong cool-core, are imaged only with ACIS-S and 
therefore we do not use them in our analysis.

\begin{figure}[h]
\begin{center}
\includegraphics[width=9.5cm, angle=0]{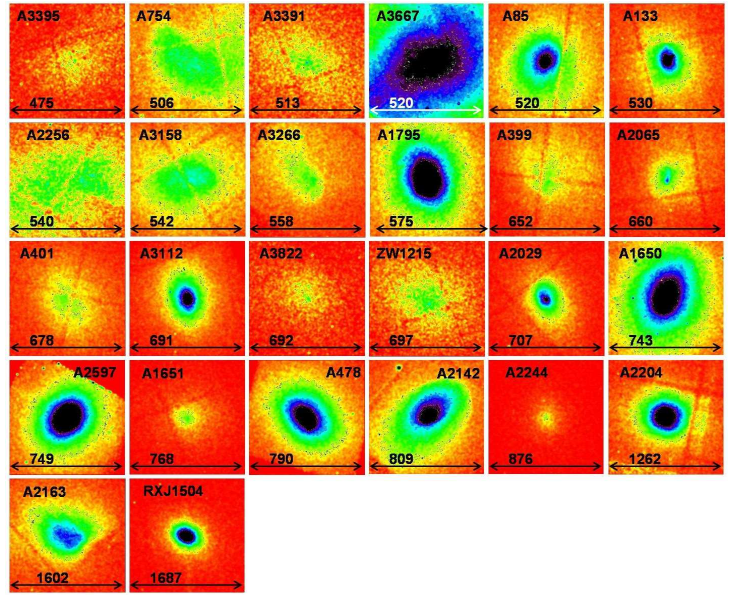}
\end{center}
\caption{Low-z cluster sample drawn from the 400 SD survey. Images are normalized by the 
exposure maps and smoothed with a gaussian kernel of 3 pixel. The images have a size of 
8$\times$8 arcmin and are ordered with increasing redshift. Physical sizes of the individual 
images are shown in kpc at the bottom of the images.}
 \label{lowz}
\end{figure}

\begin{table*}
\caption{400 SD low redshift sample: (1) cluster name, (2) Chandra observation identifier, 
(3) redshift, (4) surface brightness concentration without K-correction, (5) global temperature from Cavagnolo 
et al.  \cite{cav} (the values in parentheses are from Vikhlinin et al. \cite{vikhlinin09}), (6) core 
temperature from Cavagnolo et al. (2009), (7) soft-band luminosity in units $10^{44}$ 
[erg/s], and (8) presence of a radio source according to the NVSS archive. }  
\label{table:2}      
\centering           
\begin{tabular}{llllllll} 
\hline\hline                
Cluster  &  Obs ID & z    & $c_{SB}$ & T (keV) & T$_{core}$ & $L_{X}$  & $R_{src}$ \\   
(1)      & (2)  & (3)     & (4) &  (5) & (6)  & (7) & (8) \\
\hline                        
A3395	  & 4944  &  0.0506   & 0.081$\pm$0.006 & 5.13 (5.10) & 5.3 & 1.09 & - \\
A754	  & 577, 6796, 6797  & 0.0542   & 0.036$\pm$0.001 & 9.94 (8.73) & 8.3 & 2.78  & no \\
A3391    & 4943 & 0.0551   & 0.045$\pm$0.002 & 6.07 (5.39) & 7.0 & 1.05  & - \\
A3667	  & 513, 5751, 5752, 6296, 6292  & 0.0557   & 0.029$\pm$0.001 & 6.72 (6.33) & 5.4 & 3.14 & - \\
A85	      & 904 & 0.0557   & 0.167$\pm$0.001 & 6.90 (6.45) & 3.2 & 2.91  & yes \\
A133 	  & 2203, 9897 & 0.0569   & 0.259$\pm$0.003 & 3.71 (4.01) & 2.3 & 0.96 & yes \\	
A2256	  & 2419, 965, 1386 & 0.0581   & 0.024$\pm$0.001 & 6.90 (8.37) & 6.8 & 2.66  & no \\
A3158	  & 3201, 3712 & 0.0583   & 0.045$\pm$0.001 & 4.94 (4.67) & 5.9 & 1.72  & - \\
A3266    & 899 & 0.0602   & 0.047$\pm$0.002 & 8.35 (8.63) & 8.0 & 2.69  & - \\
A1795	  & 5289, 5290, 6159, 6161, 6162, 6163  & 0.0622   & 0.180$\pm$0.001 & 7.80 (6.14) & 3.3 & 3.52  & yes \\
A399	  & 3230 & 0.0713   & 0.040$\pm$0.001 & 5.80 (6.49) & 7.1 & 2.01  & no \\
A2065    & 3182 & 0.0723   & 0.087$\pm$0.002 & 5.35 (5.44) & 4.2 & 1.82  & yes \\
A401	  & 518, 2309 & 0.0743   & 0.040$\pm$0.001 & 8.07 (7.72) & 7.1 & 3.90  & no \\
A3112	  & 2216, 2516 & 0.0759   & 0.273$\pm$0.001 & 4.28 (5.19) & 2.7 & 2.43  & - \\
A3822 	  & 8269 & 0.0760   & 0.045$\pm$0.004 & 5.45 (5.23) & 5.6 & 1.91  & - \\
ZCl1215 & 4184 & 0.0767   & 0.040$\pm$0.002 & 6.00 (6.54) & 6.5 & 1.80  & no \\
A2029	  & 891, 4977, 6101 & 0.0779   & 0.169$\pm$0.003 & 7.38 (8.22) & 4.2 & 5.72  & yes \\
A1650 	  & 5822, 5823, 6356, 6357, 6358, 7242, 7691 & 0.0823   & 0.112$\pm$0.001 & 5.89 (5.29) & 5.1 & 2.33  & no \\
A2597 	  & 922, 6934  & 0.0830   & 0.332$\pm$0.001 & 3.58 (3.87) & 2.7 & 2.09  & yes \\
A1651	  & 4185  & 0.0853   & 0.082$\pm$0.003 & 7.00 (6.41) & 6.7 & 2.93  & yes \\
A478	  & 1669  & 0.0881   & 0.194$\pm$0.001 & 7.07 (7.96) & 3.1 & 7.24  & yes \\
A2142	  & 7692, 5005 & 0.0904   & 0.082$\pm$0.001 & 8.24 (10.04) & 6.3 & 7.20  & no \\
A2244	  & 7693 & 0.0989   & 0.103$\pm$0.005 & 5.57 (5.37) & 5.4 & 2.98  & no \\
A2204    & 6104, 7940 & 0.1511   & 0.315$\pm$0.002 & 6.97 (8.55) & 3.5 & 9.35  & yes \\
A2163	  & 1653 & 0.2030   & 0.025$\pm$0.002 & 12.12 (14.72) & 16 & 13.7  & no  \\
RXC J1504.1-0248 & 5793 & 0.2169   & 0.333$\pm$0.003 & 8.90 (9.89) & 4.6 & 15.6  & yes \\
Hydra*    & 4969, 4970 & 0.0549   &  - & 4.30 (3.64) & 2.9 & 1.93  & yes \\
S1101*   &  1668 & 0.0564   & -  & 2.65 (2.44) & 2.0 & 1.03 & - \\
\hline  

\end{tabular}
\begin{list}{}{}
\item$^{\mathrm*}$ Strong cool-core clusters observed only with ACIS-S which we did not consider
in our analysis
\end{list}
\end{table*}

\section{The distant cluster samples}

Beyond redshift 0.5 there are only three X-ray complete cluster samples, all 
selected from ROSAT PSPC pointed observations.  
They  are: (i) the 400 SD (Burenin et al. \cite{burenin}) high-$z$ sample which include all 
clusters from the 400 SD catalog with $z\ge 0.5$; (ii) the Rosat Deep Cluster Survey (RDCS, 
Rosati et al. \cite{rosati98}); and (iii) the Wide Angle ROSAT Pointed Survey (WARPS, 
Jones et al. \cite{jones}).  

These samples are still limited by small number statistics.  Furthermore, observations of 
distant clusters undergo a strong surface brightness dimming $\propto (1+z)^{-4}$ and 
have a small angular size, thus the study of their central regions requires the 
high-resolution provided only by \textit{Chandra}.  Its excellent spatial resolution 
is mandatory to measure the central surface brightness of high redshift clusters, to the point
that even the use of XMM-Newton, with a much larger collecting area but with a poorer 
resolution, would not be suitable for our study.
While the distant 400 SD sample has been fully observed with {\sl Chandra}, the RDCS and WARPS 
samples have been only partially observed with a {\sl Chandra} follow up.  For this 
reason, we merge them into the RDCS+WARPS sample.  We point out that the RDCS+WARPS samples 
reach lower fluxes in comparison with the 400 SD sample, and therefore include the highest
redshift clusters.

Current X-ray surveys from serendipitous pointings, such as the XMM-LSS (Pierre et al. \cite{pierre}), 
the XMM-Newton Cluster Survey (Sahl{\'e}n et al. \cite{xcs}), the XMM-Newton Distant Cluster 
Project (XDCP\footnote{http://www.xray.mpe.mpg.de/theorie/cluster/XDCP/xdcp\_index.html}, B\"ohringer et al. \cite{boehringer05}), and the Swift X-ray Cluster
Survey (Moretti et al. 2010, in prep.) will add new $z\sim 1 $ X-ray clusters in 
the next years.  However, new detections will need an appropriate {\sl Chandra} follow-up to 
have a better spatial resolution (in case of clusters detected with other instruments) or a 
better signal-to-noise ratio (in case of cluster discovered in shallow {\sl Chandra} images).  
Another fundamental aspect is that in order to derive meaningful constraints on the evolution 
of high-$z$ clusters, we must use samples with a well defined selection function.  The three
high redshift samples used in this work are consistent with each other in terms of number 
counts (see Burenin et al. \cite{burenin}), hence they do not have evident bias in their 
selection function.

\subsection{The 400 SD high-$z$ sample: 0.5 $<$ z $<$ 0.9}

This sample (see Table 2, Fig. 2) includes 20 clusters with $0.5 < z < 0.9$. 
This is a subsample of the list of clusters used in Vikhlinin et al. (\cite{vikhlinin09b}) to 
constrain cosmological parameters.  We opted not to use the same sample (starting 
at $z = 0.35$) to avoid evolutionary effects within the sample, and in order to have 
an average redshift comparable to that of the RDCS+WARPS sample.  All clusters 
have been observed with {\sl Chandra} with an exposure time long enough to gather at 
least 2000 net counts for each cluster. Therefore the quality of these images is suitable 
for our study.

\begin{table*}
\caption{400 SD high z sample: (1) cluster name, (2) Chandra observation identifier, (3) redshift, (4) spectral temperature, (5) total rest-frame soft-band luminosity in units $10^{44}$ 
[erg/s], (6) surface brightness concentration, and (7) presence of a radio source in the cluster center according to NVSS.}             
\label{table:2}      
\centering           
\begin{tabular}{lllllll} 
\hline\hline                
Cluster  & Obs ID & z & T (keV) & $L_{X}$  & $c_{SB}$ & $R_{src}$ \\   
(1)      & (2) & (3)   & (4) & (5) & (6) & (7)  \\
\hline                       

WARP J0030.5+2618	& 5762  &  0.500    & $5.62^{+1.53}_{-1.04}$  & 1.65 & 0.040$\pm$0.011 & no \\
400d J1002+6858	    & 5773 &  0.500    & $6.86^{+1.01}_{-0.72}$  & 2.41 & 0.060$\pm$0.012 & yes \\
WARP J1524.6+0957	& 1664 &  0.516    & $4.59^{+0.65}_{-0.40}$  & 2.29 & 0.032$\pm$0.006  & no \\
400d J1357+6232	    & 5763,  7267   &  0.525    & $5.29^{+0.61}_{-0.56}$  & 1.87 & 0.054$\pm$0.010  & no  \\
400d J1354-0221        & 4932, 5835  &  0.546    & $6.82^{+1.59}_{-1.13}$  & 4.57 & 0.043$\pm$0.009  & no \\
400d J1117+1744       & 4933, 5836  &  0.547    & $5.99^{+1.41}_{-1.45}$  & 0.78 & 0.041$\pm$0.010 & no  \\
400d J1120+2326	   & 3235  &  0.562    & $5.80^{+0.79}_{-0.54}$  & 5.06 & 0.027$\pm$0.011  & no \\
WARP J0216.5-1747    & 5760, 6393 &  0.578    & $5.95^{+0.72}_{-0.72}$  & 1.01 & 0.055$\pm$0.014  & no \\
400d J0521-2530 	     &  5758 &  0.581    & $2.98^{+1.51}_{-0.91}$  & 0.51 & 0.046$\pm$0.007  & no \\
400d J0956+4107        &  5294, 5759  &  0.587    & $6.86^{+1.01}_{-0.72}$  & 2.01 & 0.040$\pm$0.007  & no \\
400d J0328-2140	     & 5755, 6258 &  0.590    & $5.54^{+0.61}_{-0.48}$  & 2.47 & 0.062$\pm$0.009  & no \\
WARP J1120.1+4318* 	 & 5771  &  0.600    & $4.68^{+0.53}_{-0.46}$  & 4.37 & 0.063$\pm$0.005  & no \\
ZwCl 1332.8+5043       & 5772  &  0.620    & $5.14^{+1.26}_{-1.10}$  & 2.42 & 0.068$\pm$0.017  & no \\
RDCS J0542-4100* 	     & 914  &  0.642    & $6.63^{+0.83}_{-0.73}$  & 3.55 & 0.043$\pm$0.007  & no \\
400d J1202+5751  	     &  5757  &  0.677    & $5.96^{+1.35}_{-0.91}$  & 2.11 & 0.030$\pm$0.008  & no \\
400d J0405-4100 	     & 7191 &  0.686    & $3.28^{+0.68}_{-0.64}$  &7.78  & 0.073$\pm$0.009  & no \\
400d J1221+4918  	     & 1662 &  0.700    &	$7.45^{+1.04}_{-0.64}$  & 3.60 & 0.026$\pm$0.006  & yes \\
400d J0230+1836   	     & 5754 &  0.799    & $5.78^{+1.23}_{-0.74}$  & 3.28 & 0.036$\pm$0.009  & no \\
WARP J0152.7-1357N*	 & 913  &  0.833    & $5.30^{+1.05}_{-0.78}$  & 3.41 & 0.027$\pm$0.008  & no \\
WARP J1226.9+3332*	 & 3180, 5014 &  0.890    & $12.34^{+1.20}_{-0.95}$ & 9.68 & 0.086$\pm$0.007  & yes \\
\hline  

\end{tabular}
\begin{list}{}{}
\item$^{\mathrm*}$ also in the RDCS+WARPS
\end{list}
\end{table*}

\begin{figure}[h]
\begin{center}
\includegraphics[width=9.2cm,angle=0]{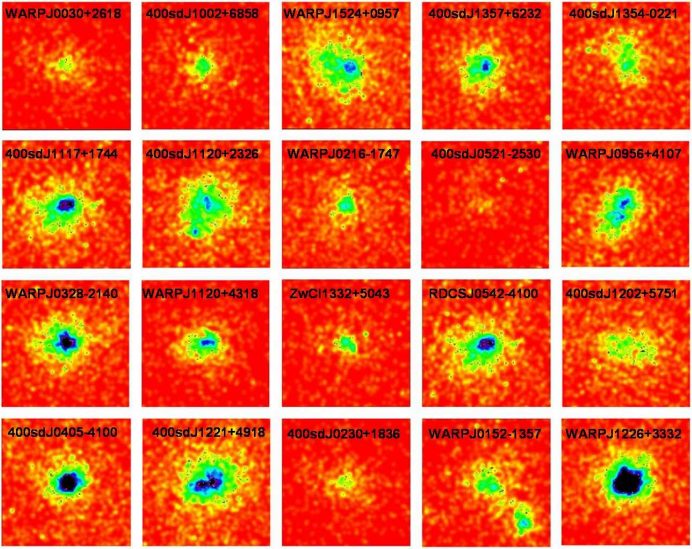}
\end{center}
\caption{400 SD high-z sample. Images are normalized by the exposure maps and smoothed with a gaussian 
kernel of 3 pixel. The images have a size of 4$\times$4 arcmin and are ordered with increasing redshift 
 (from top left to bottom right). }
 \label{400high}
\end{figure}

\subsection{The RDCS+WARPS sample: 0.6 $<$ z $<$ 1.3}

The RDCS has provided a deep, large area, X$-$ray selected cluster sample, with 115 spectroscopically-confirmed X-ray clusters of galaxies at $S> 3 \times 10^{-14}$ erg 
s$^{-1}$ cm$^{2}$ in the soft-band, selected from an area of 50 deg$^{2}$, in a 
homogeneous and objective manner via a serendipitous search in ROSAT PSPC deep
pointings.  A deeper subsample with $S > 1\times 10^{14}$ erg s$^{-1}$ cm$^{2}$ is 
defined over a more limited area of 5 deg$^{2}$.  To date, this subsample has the 
greatest number of spectroscopically identified distant clusters, with 11 at 
$z > 0.6$ for $S > 3 \times 10^{-14}$ erg s$^{-1}$ cm$^{2}$, and 17 for $S > 1 \times 
10^{-14}$ erg s$^{-1}$ cm$^{2}$.

The WARPS is an X-ray selected survey for high redshift galaxy clusters 
based on serendipitous detections in targeted ROSAT PSPC observations. The survey covers 
an area of 71 deg$^{2}$ and contains a complete sample of 129 clusters with a flux limit of 
$S \sim 6.5 \times 10^{-14}$ erg s$^{-1}$ cm$^{2}$.

\begin{figure}[h]
\begin{center}
\includegraphics[width=9.2cm,angle=0]{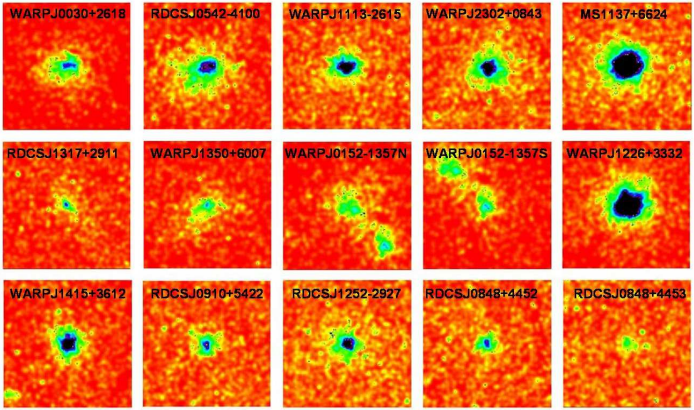}
\end{center}
\caption{RDCS+WARPS sample. Images are normalized by the exposure maps and smoothed 
with a gaussian kernel of 3 pixel. The individual images have a size of 4'$\times$4' and are 
ordered with increasing redshift (from top left to bottom right).}
 \label{rdcs}
\end{figure}

The merged sample (see Table 3, Fig.~\ref{rdcs}) includes all clusters 
from the RDCS with redshift greater than 0.6 (10 out of 15 objects) observed with 
\textit{Chandra}, in addition to 7 (out of 12) WARPS clusters with similar data. 
There is an overlap of 2 objects between the two samples 
(RDCS J0542-4100, WARP J0152.7-1357N), hence the combined sample has a total of 15 clusters.
We note that four clusters are common to the RDCS+WARPS and the 400 SD high-$z$: 
WARP J1120.1+4318, RDCS J0542-4100, WARP J0152.7-1357N and WARP J1226.9+3332.

\begin{table*}
\caption{RDCS+WARPS sample: (1) cluster name, (2) Chandra observation identifier, (3) 
redshift, (4) spectral temperature, (5) total rest-frame soft-band luminosity in units $10^{44}$
 [erg/s]), (6) surface brightness concentration without K-correction, and (7) presence of a radio 
 source in the cluster center according to NVSS.}
    
\label{table:2}      
\centering           
\begin{tabular}{lllllll} 
\hline\hline                
Cluster & Obs ID & z    & T (keV)     &  $L_{X}$   & $c_{SB}$  & $R_{src}$ \\   
(1)     & (2)  & (3)         & (4)         & (5)      & (6)  & (7) \\
\hline                       
WARP J1120.1+4318	& 5771 & 0.600     & $4.67^{+0.53}_{-0.45}$  & 4.77 & 0.063$\pm$0.007 & no \\
RDCS J0542-4100  	    & 914  & 0.642     & $6.63^{+0.83}_{-0.73}$  & 3.55 & 0.043$\pm$0.007 & no \\
WARP J1113.0-2615 	&  915 & 0.730     & $5.46^{+0.68}_{-0.55}$  & 1.51 & 0.088$\pm$0.014 & no \\
WARP J2302.8+0843	& 918  & 0.734     & $5.32^{+0.60}_{-0.54}$  & 1.73 & 0.061$\pm$0.009 & no \\
MS 1137.5+6624         & 536 & 0.782      & $6.72^{+0.46}_{-0.45}$  & 4.05 & 0.097$\pm$0.007 & no \\
RDCS J1317+2911       &	2228  & 0.805   & $6.58^{+2.95}_{-2.02}$ & 0.58  & 0.108$\pm$0.022  & no \\
WARP J1350.8+6007 	& 2229 & 0.810     & $3.96^{+0.66}_{-0.45}$ & 2.73  & 0.064$\pm$0.015  & no \\
WARP J0152.7-1357N  & 913 & 0.828      & $5.14^{+0.86}_{-0.65}$ & 3.41  & 0.027$\pm$0.08  & no \\
WARP J0152.7-1357S  & 913 & 0.835      & $5.30^{+1.05}_{-0.78}$ & 2.04  & 0.051$\pm$0.014  & no \\
WARP J1226.9+3332	& 3180, 5014        & 0.890     & $12.34^{+1.20}_{-0.95}$ & 9.68 & 0.086$\pm$0.07  & yes \\
WARP J1415.1+3612	& 4163    & 1.030    & $5.91^{+0.71}_{-0.61}$ & 4.12  & 0.144$\pm$0.016  & yes \\
RDCS J0910+5422	   & 2227, 2452 & 1.106     & $5.48^{+1.79}_{-1.23}$ & 0.79  & 0.101$\pm$0.021  & no \\
RDCS J1252-2927    	& 4198, 4403 & 1.235     & $6.45^{+1.08}_{-0.96}$ & 2.34  & 0.082$\pm$0.014  & no \\
RDCS J0848.9+4452 (Lynx E)   & 927, 1708  & 1.261     & $4.94^{+1.70}_{-1.04}$ & 0.82  & 0.089$\pm$0.022  & no \\
RDCS J0848.6+4453 (Lynx W)   &  927, 1708  & 1.273     & $2.11^{+1.22}_{-0.54}$ & 0.34  & 0.080$\pm$0.032  & no \\
\hline  

\end{tabular}
\end{table*}

\section{Data reduction and spectral analysis}
 All data was taken from the public \textit{Chandra} archive.  We
performed a standard data reduction starting from the level=1 event
files, using the {\tt CIAO 4.1} software package, with a recent
version of the Calibration Database ({\tt CALDB 4.2}).  For
observations taken in the VFAINT mode we ran the task {\tt
  acis$\_$process$\_$events} to flag background events that are most
likely associated with cosmic rays and distinguish them from real
X-ray events. With this procedure, the ACIS particle background can be
significantly reduced compared to the standard grade selection.  We
also apply the CTI correction to the observations taken when the
temperature of the Focal Plane was 153 K.  This procedure allows to
recover the original spectral resolution partially lost because of the
CTI. The correction applies only to ACIS-I chips, since the ACIS-S3
did not suffer from radiation damage.  The data is filtered to include
only the standard event grades 0, 2, 3, 4 and 6. We checked visually
for hot columns left after the standard reduction.  We identify the
flickering pixels as the pixels with more than two events contiguous
in time, where a single time interval was set to 3.3~s. For exposures
taken in VFAINT mode, there are practically no flickering pixels left
after filtering out bad events. We finally filter time intervals with
high background by performing a $3\sigma$ clipping of the background
level using the script {\tt analyse\_ltcrv}.  The removed time intervals
always amount to less than 5\% of the nominal exposure time for ACIS-I
chips. We remark that our spectral analysis is not affected by any
residual flare, since we are always able to compute the background
from source-free regions around the clusters from the same observation
(see below), thus taking into account any possible spectral distortion
of the background itself induced by the flares.

For the clusters in the distant samples, we perform a simple spectral
analysis extracting the spectrum of each source from a circular region
with radius $R_{ext} = 40$\arcsec centered around the X-ray centroid.
The background is always obtained from empty regions of the chip in
which the source is located. This is possible since all the distant
clusters have an extension of less than 3 arcmin (corresponding to 
a range in physical size of 1100-1500 kpc), as opposed to the 8
arcmin size of the ACIS-I/-S chips. The background file is scaled to
the source file by the ratio of the geometrical area. In principle, the
background regions may partially overlap with the outer virialized
regions of the clusters. However, the cluster emission from these
regions is negligible with respect to the instrumental background, and
does not affect our results. Our background subtraction procedure, on
the other hand, has the advantage of providing the best estimate of
the background for that specific observation.  The response matrices
and the ancillary response matrices of each spectrum are computed
respectively with {\tt mkacisrmf} and {\tt mkwarf}, for the same
regions from which the spectra are extracted.

The spectra are analysed with XSPEC v12.5 (Arnaud et al. \cite{arnaud}) and fitted
with a single-temperature {\tt mekal} model (Kaastra \cite{kaastra}; Liedahl et al. \cite{lied}) 
using the solar abundance of Grevesse \& Sauval (1998).  The
fits are performed over the energy range $0.5-8.0$ keV.  The free
parameters in our spectral fits are temperature, metallicity and
normalization.  Local absorption is fixed to the Galactic neutral
hydrogen column density ($N_H$) taken from Wilms et al. (\cite{wilms}).  We
used Cash statistics applied to the source plus background, which is
preferable for low S/N spectra (Nousek \& Shue \cite{nousek}).  Redshifts are
known from the literature for all the sources.  The X--ray luminosity
is computed in the rest-frame soft-band after extrapolating the
surface brightness up to 1 Mpc (see Section 6).  Best-fit temperatures
and luminosities as shown in Table 2 and 3.  
We also attempted to detect
temperature gradients by dividing the cluster emission into an inner
and an outer region, roughly with the same number of net counts.
However, the difference between the two best fit values are generally
much smaller than the one sigma error bars, hampering any meaningful
conclusion on the presence of gradients in the high-redshift sample.
We ascribe this result to the systematic bias towards higher
temperatures when the signal-to-noise ratio is degraded in thermal
spectra (see Leccardi \& Molendi \cite{leccardi07}).  This effect 
tends to cancel the temperature gradients
expected in cool-core clusters, making it impossible to characterize
high redshift cool-cores on the basis of a spectral analysis.

For the local sample, a spatially resolved spectral
analysis is in principle possible, given the much larger 
signal-to-noise. However, for simplicity we rely on the 
temperature and luminosity values
measured by Cavagnolo et al. (\cite{cav}) and Vikhlinin et al. (\cite{vikhlinin09}) (see Table 1).

\section{Surface brightness concentration, $c_{SB}$ }

The simplest observational signature of the presence of a cool-core is
a central spike in the surface brightness profile.  This is also the
only possible diagnostic we can apply to high redshift clusters given
the difficulty in performing spectral analysis to detect the
temperature decrease in the core region.  There are two ways to
evaluate the central spike: fitting a surface brightness profile and
measuring the inner slope, or measuring the integrated emission within
a given radius.  The first method requires high signal-to-noise 
and suffers from systematics due to the choice of the fitting model.  
The second approach, which relies on integrated quantities, has the great 
advantage to be robust and minimizes the noise.

In Santos et al. (\cite{joana}) we defined the phenomenological parameter
$c_{SB}$ that quantifies the excess emission in a cluster core by
measuring the ratio of the surface brightness within a radius of 40
kpc with respect to the SB within a radius of 400 kpc: $c_{SB} =
SB (r<40 kpc) / SB (r<400 kpc)$. We note that using the slightly 
different definition $c_{SB} =SB (r<40 kpc) / SB (40<r<400 kpc)$, does not
significantly improve the sensitivity of this estimator.
This simple parameter has been shown to be robust and particularly 
useful when dealing with the low S/N data of distant clusters (e.g., Hudson et al.\cite{hudson}). 
Instead, the cuspiness parameter (Vikhlinin et al. \cite{vikhlinin06}) is based on the central density slope, 
which is more sensitive to the details in the SB profile. 
Given the range of signal-to-noise of our objects, the results based on the measured 
slope of the inner SB may be affected by large variance in the best fitting values.  
On the other hand, our parameter $c_{SB} $ is based on simple photometry, and it provides an 
homogeneous characterization of our cluster sample which spans a wide range of S/N.

We validated the redshift independence of $c_{SB}$
(apart from possible K-corrections as described in the next
subsection) by cloning low-$z$ clusters to high redshift, therefore
$c_{SB}$ can be directly applied to the distant samples.
In addition, we also showed that $c_{SB}$ strongly anti-correlates
with the central cooling time.  Therefore we use this parameter to
characterize the cluster population at high redshift, and, in order 
to have a fair comparison, we also apply 
the same diagnostics to the local clusters.
We stress that working with a physical size instead of a 
scaled radius for the cooling region is crucial for this study. In Santos et al. (\cite{joana}) we 
discussed at length this aspect and we concluded that the cooling 
radius does not evolve significantly with redshift, which is expected 
since the cool core phenomena is related with non-gravitational physics.

In this analysis we use the \textit{Chandra} images in the soft-band
(0.5-2.0 keV), after accounting for the vignetting effect computed
at 1.5 keV.  This is achieved by normalizing the exposure maps to the
total exposure time and dividing the images by the normalized
maps. This correction is particularly relevant for the measurement of
$c_{SB}$ in the local clusters, given their larger extension which
implies a larger vignetting correction.  With this procedure we also
automatically correct for the gaps between the detector chips.  In a
few cases (A478, A2597, observed only with ACIS-S, and A85, whose
observation lies close to the edge of one of the ACIS-I  chips), the 400 kpc
radius lies beyond the edge of the observed field.  In these cases the
measurement of $c_{SB}$ requires an extrapolation of the surface
brightness from about 300 kpc to 400 kpc according to the best fit
$\beta$-model, as described in Section 6.

\subsection{K-correction applied to the $c_{SB}$ parameter: ``beheading''}

A cluster with a single temperature ICM will be observed with the
same value of $c_{SB}$ at any redshift.  However, in the presence of a
cool-core, the spectral emission from the central region would have a
softer spectrum with respect to the higher-temperature emission from
the outer regions.  This would introduce a different amount of
K-correction between the inner 40 kpc emission and the total 400 kpc
emission.  For distant clusters it is impossible to derive the
rest-frame $c_{SB}$ value, since it would require the knowledge of the
temperature profile of the ICM.  However, in order to compare the
$c_{SB}$ distribution of local and distant clusters, we can follow the
alternative approach of applying the K-correction to local clusters as
if they were at the average redshift of the distance cluster sample.
This is possible since we can exploit the knowledge of the temperature
profiles in local clusters. We apply the K-correction in the assumption
 of no evolution of the temperature and metallicity profile, in order to test 
 the null hypothesis of no evolution in the CC population.

We proceed as follows.  We select a subsample of local clusters with a
wide range of cool-core strength, with a signal-to-noise ratio large
enough to allow us to perform spatially resolved spectroscopy.  The
clusters are listed in Table \ref{clusterlist}.  All the clusters are
observed with ACIS-I, and the radius corresponding to 400 kpc is
always within the observed field.  We divide the clusters into rings
with equal number of net detected counts (about 3000 in the soft-band)
and we measure the projected temperature and metallicity\footnote{We
  are not interested in deprojected quantities here, since we just
  want to estimate the K-correction effect on the surface brightness}.
In Table \ref{clusterlist} we also show the minimum and maximum
temperatures measured in each cluster.  After freezing the best fit
parameters in each ring, we simulate the expected net counts for the
same cluster at different redshifts, including therefore the most
accurate K-correction.  From the cloned surface brightness profile we
measure the $c_{SB}$ values at different redshifts.  In this way we
are able to quantify the apparent evolution of $c_{SB}$ as a function
of redshift for the same cluster.  The effect of this apparent
evolution, uniquely due to the X-ray K-correction, must be removed
when comparing local and distant clusters.

\begin{table}
\caption{Clusters used to compute the effect of
  K-correction on $c_{SB}$. \label{clusterlist} }
\begin{tabular}{llllll}
Cluster  & z & kT (keV)  & $N_H/10^{22}$ cm$^{-2}$  & $c_{SB}(0)$\\
\hline
\hline
RXJ1504 & $0.229$ & $4.4-9.0$ &  $0.060$  & $0.343$ \\
A1835 & $0.25$ & $3.8-10.0$ &  $0.023$  & $0.258$ \\
ZW3146 & 0.291 & $4.0-9.0$ &  $0.03$  & $0.210$ \\
A907  &  $0.153$ &  $3.5-7.0$ &  $0.054$  & $0.164$ \\
A2029 &  $ 0.078$ &  $5.3-8.5$  & $0.030$  & $0.16$ \\
A2163 & $0.203$  & 12.0-16.0  & $0.012$ &  $0.044$ \\
A3266 & $0.0602$ & 7.5-11.0  & $0.0162$ &  $0.059$ \\
A2597 & $0.0803$ & 2.1-4.0 & 0.$025$  & $ 0.363$\\
\hline
\end{tabular}
\end{table}

\begin{figure}
\includegraphics[width=8.cm,angle=0]{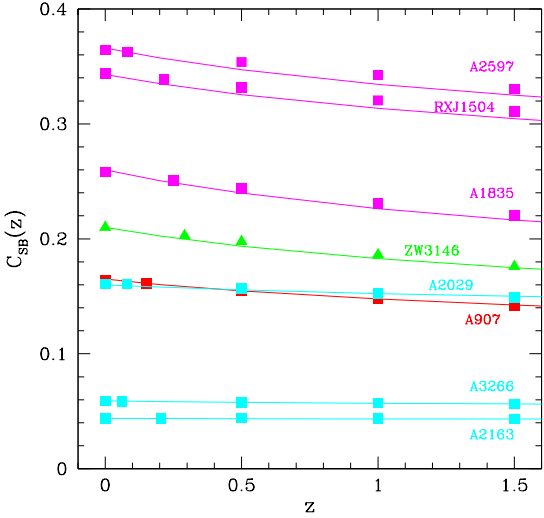}
\caption{$c_{SB}(z)$ for the clusters in Table \ref{clusterlist}.
  Continuous lines are approximate fits with a power law
  $\propto (1+z)^{-\alpha}$.}
\label{csbz}
\end{figure}

\begin{figure}
\includegraphics[width=8cm,angle=0]{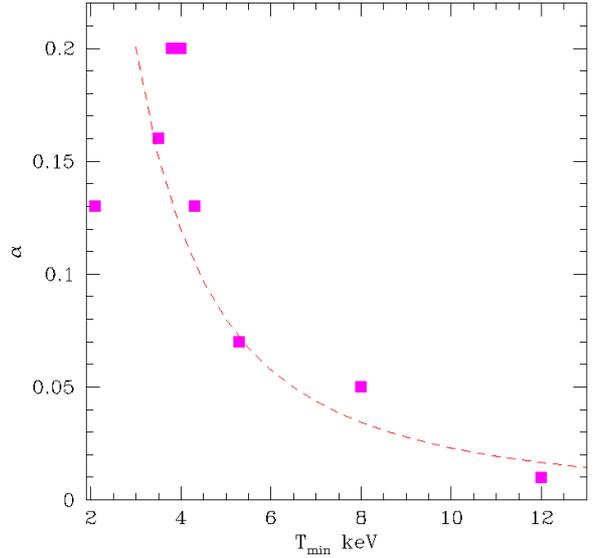}
\caption{Exponent $\alpha$ vs minimum (central) temperature for clusters in Table 4. The dashed line represents eq. (1).}
\label{alpha}
\end{figure}

We approximate the apparent evolution of the $c_{SB}$ parameter as a
function of redshift for this set of clusters.  As shown in Figure
\ref{csbz}, the average behavior of $c_{SB}(z)$ for cool-core
clusters can be described with a simple power law.  For strong
cool-core clusters, we have roughly $\propto (1+z)^{-\alpha}$, with
$\alpha \sim 0.15$, which implies a decrease of about $13\%$ at $z\sim
1$.  On the other hand, for weaker cool-cores (lower $c_{SB}$) we
expect a slower evolution.  The K-correction is null for a flat
temperature profile.  We indeed find that $\alpha$ correlates with the
minimum temperature, as shown in Figure \ref{alpha}.  

In order to predict an unbiased $c_{SB}$ distribution at high-$z$ we
need to apply a correction of the order $\leq 10\%$ as a function of
redshift and cool-core strength (or minimum central temperature).  A
simple approximation, shown by the dashed line in Figure \ref{alpha}
reads:

\begin{equation}
\alpha = 0.08 \times (kT_{min}/5)^{-1.8}\, ,
\end{equation}

\noindent with a maximum value $\alpha = 0.2$.  Therefore, we can modify the
$c_{SB}(0)$ values of the local sample by a factor $\propto
(1+z)^{-\alpha(T_{min})}$, considering the median redshift of the 
RDCS+WARPS sample, $<z>=0.83$.  This correction corresponds to an average 
decrease in  $c_{SB}$ by 4.3\%, relative to the original values, 
and reaches at most 20\% for the low temperature strong cool-cores A133 and A2597.  
The modified distribution can now be directly compared with that of 
distant clusters.

We also checked that the different values of the Galactic hydrogen column density, 
both in the local and in the high-$z$ samples, do not significantly affect the measured $c_{SB}$. 
The corrections are expected to be below 1-2\%, and therefore we neglect this small effect.

\subsection{The $c_{SB}$ distributions}

\begin{figure*}
\begin{center}
\includegraphics[width=9cm,angle=0]{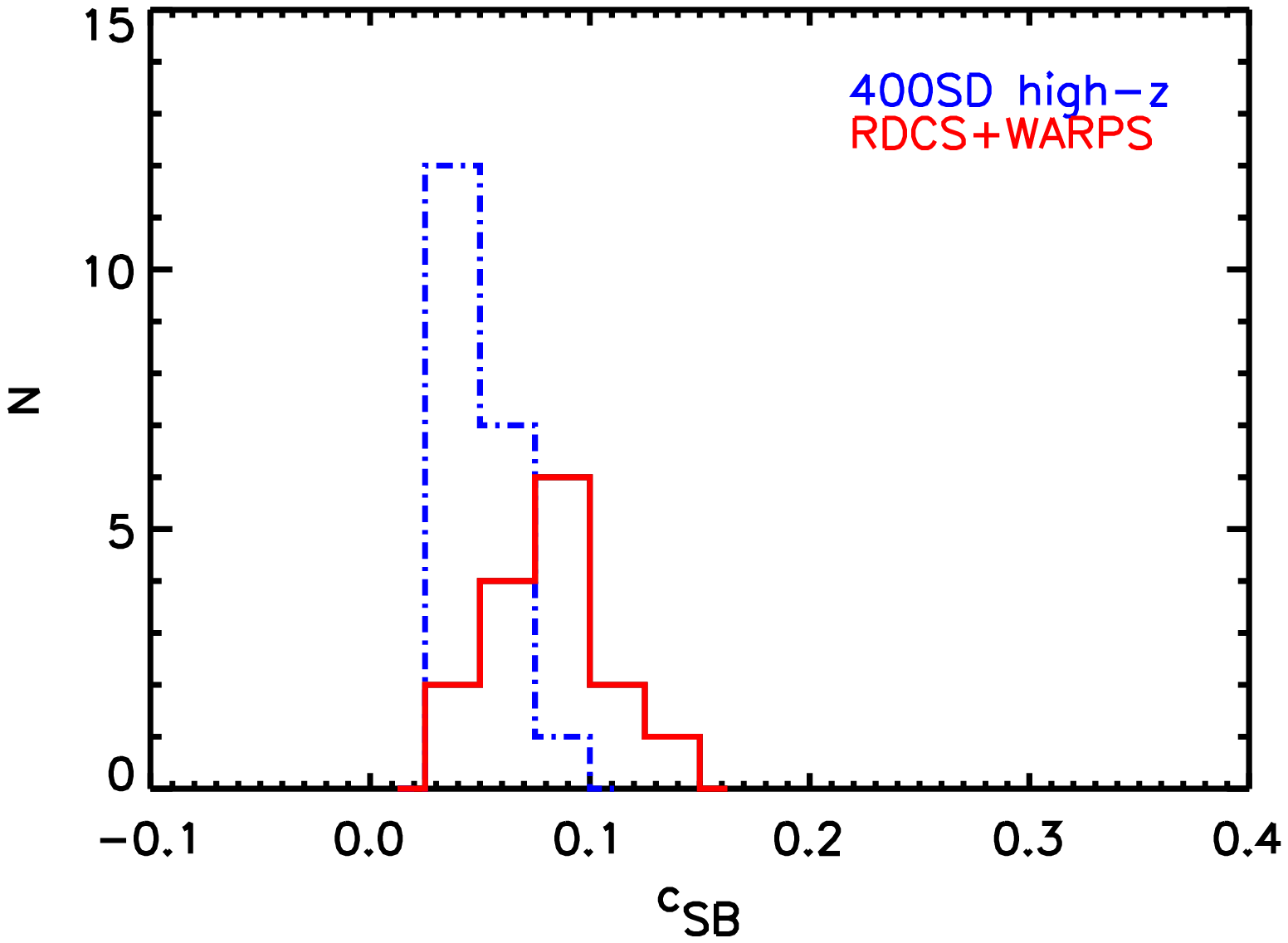}
\includegraphics[width=9cm,angle=0]{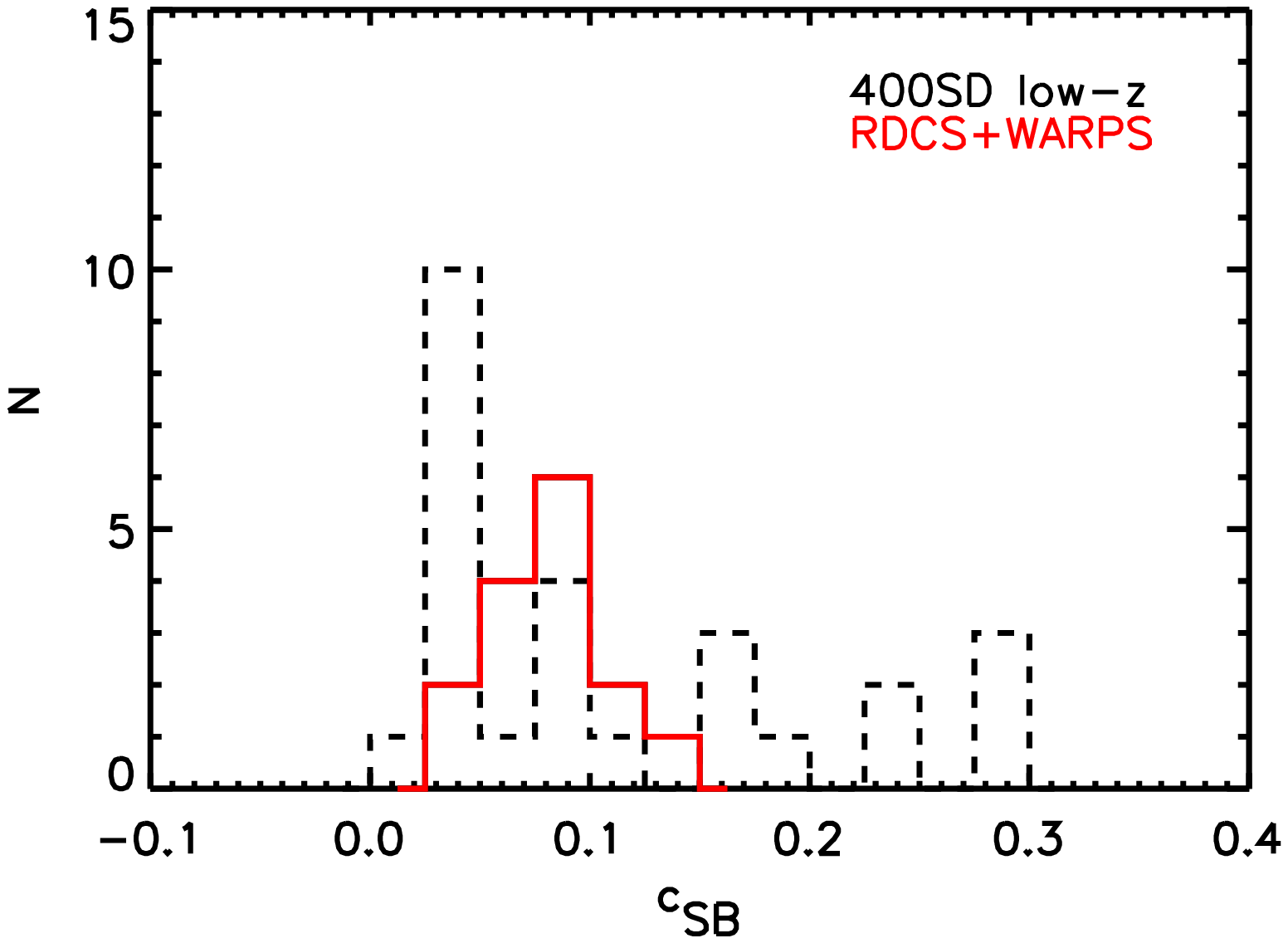}
\end{center}
\caption{\textit{Left} Distributions of $c_{SB}$ for the two distant 
samples, RDCS+WARPS (solid, red) and 400 SD high-$z$ (dash-dot, blue), and \textit{right} 
comparison between the K-corrected $c_{SB}$ distribution of the local (dash, black) and the RDCS+WARPS samples. }
 \label{Figcsb}
\end{figure*}

Before comparing the $c_{SB}$ distribution of the local and distant
samples, we compare the two distant samples separately.  Their
$c_{SB}$ distributions are shown in Figure \ref{Figcsb}, left panel.
Quite unexpectedly, the shape and range of the two high-$z$ $c_{SB}$
distributions are statistically different.  We perform a K-S test and
 find a null hypothesis probability of $0.6$\%, implying that the two distant 
 samples do have different distributions of cool-core strength.
Taking into account that there are four clusters common to both samples, we 
repeated the test twice removing the common clusters from each sample and 
assigned them to the other one. This does not significantly affect the result of the 
K-S test, since we obtain null probabilities of $0.1$\% and $0.3$\%. 
The 400 SD high-$z$ reaches $c_{SB}$ = 0.10, with median $c_{SB}=0.043$,
whereas the RDCS+WARPS reaches $c_{SB}=0.15$, with a median $c_{SB}$ value
 equal to 0.082.  The RDCS+WARPS clusters have thus a
significantly higher surface brightness concentration with respect to
the 400 SD clusters, as can be appreciated by a visual inspection of
the cluster physical morphologies shown in Figures 2 and 3. 
The difference in the surface brightness concentration between the two
samples is confirmed by the $\beta$-model fitting performed in Section
6.  These results point to a smaller dynamical range in surface brightness 
in the 400 SD high-$z$, relative to the RDCS+WARPS.
We can exclude significant effects from K-correction or intrinsic
evolution, given the large redshift overlap between the two samples,
even though the RDCS+WARPS sample reaches a higher redshift ($z$=1.27)
with respect to the 400 SD high-$z$. Both the RDCS+WARPS and the 400 SD 
are samples based on ROSAT data, but are built with different selection 
criteria. 

We argue that this difference is likely due to a bias of the detection 
algorithm used in the 400SD survey against compact clusters with a relatively 
high mean surface brightness. This does not lead necessarily to sample incompleteness, 
as a different detection sensitivity for varying $\beta$ and core radii is taken into 
account when computing the selection function and the corresponding survey volume. 
This likely explains why number counts and luminosity functions of the three surveys 
are in very good agreement with each other. 
To check for these effects, we need to perform a detailed comparison of 
the selection criteria in the three surveys. This task goes beyond the 
scope of this work. In order to investigate evolution in the cool-cluster 
population, we decide to use the RDCS+WARPS only.

The K-corrected $c_{SB}$ distribution of the local sample (shown on the right
panel of Figure \ref{Figcsb}) spans a broad range of values and
reaches $c_{SB}$=0.315, with a significant peak at low $c_{SB}$ ($\sim 0.04$) and a median 
$c_{SB}$ equal to 0.079. 
We do not find any hint of a cool-core / non cool-core bimodality, as sometimes 
claimed in the literature (e.g., Sanderson et al. \cite{sanderson}), however we believe that 
higher statistics are needed to check for this effect.

We performed a K-S test comparing the local and distant RDCS+WARPS samples 
and found a null hypothesis probability of $p= 16$\%, 
showing that the two  $c_{SB}$ distributions are comparable.
Therefore, in our analysis we find that the bulk of the populations of
distant and local clusters have comparable cool-core strength, despite the absence 
of strong cool-core clusters at high-$z$.   
Qualitatively, our findings are compatible with a
significant population of cool-core clusters already at redshift
$z\sim 1.2$ (5 Gyr after the Big Bang), while strong cool-cores must
wait for a longer time span before they can develop (9 Gyr if they
appear below $z\sim 0.5$).  To reinforce these constraints, it is
necessary to use larger samples of high-$z$ clusters and to attempt a
better characterization of distant cool-cores.  The first occurrence
is strongly limited by the lack of wide area cluster survey in the
{\sl Chandra} and {\sl XMM-Newton era}, while the second strategy
would require the use of a significant amount of {\sl Chandra} time.
It would be worth pursuing at best these two strategies, since the
alternative is waiting for the next generation wide-area X-ray surveys
with high spatial resolution.

For completeness we also report the results obtained using the local and 
distant 400 SD samples. The K-S test provides a probability of $0.5$\% that
the two samples are drawn from the same distribution.  This would be
in agreement with the claim of strong cool-core evolution made by
Vikhlinin et al. (\cite{vikhlinin06}).  However we remark that such result, at odds
with our findings, is unlikely to be a fair description of the cluster
population at redshift greater than 0.5.

We assessed the impact of the 10 missing clusters from the RDCS+WARPS sample, i.e. clusters
which have not been observed with \textit{Chandra}, in our results.  
Assuming that all missing clusters are non-CC, the most conservative estimate of the 
fraction of CC in the RDCS+WARPS sample is 30\%, which is significantly higher than the 
10\% fraction in the 400 SD high-z samples.  
 Hence, we conclude that our results are not substantially affected by this incompleteness.

\section{Surface brightness profile fitting }

In this Section we derive a detailed description of the surface
brightness profiles of the clusters in our sample, which will be
later used to compute the gas density profiles, the entropy and 
the central cooling times.  
Unless one has very precise data, the isothermal $\beta$-model proposed by 
Cavaliere \& Fusco-Femiano (\cite{cavaliere}) provides a fair description of 
the radial X-ray surface brightness profiles of galaxy clusters.
We fit the azymuthally averaged profiles and use the fit parameters to characterize 
the morphology of the distant clusters.

\begin{figure}[h]
\begin{center}
\includegraphics[width=8.5cm,angle=0]{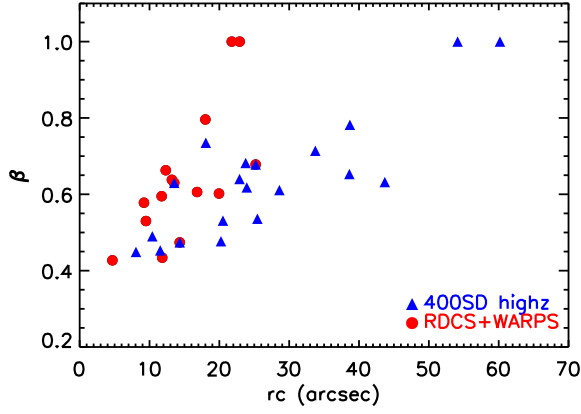}
\end{center}
\caption{Correlation between the single $\beta$-model parameters,
  $r_{c}$ and $\beta$ for the two samples of high-$z$ clusters.}
 \label{Figbeta}
\end{figure}

Distant clusters have a small angular size and low photon statistics,
therefore a single-$\beta$-model provides a good fit to the
radial profiles.  Our attempts to fit a double $\beta$-models to
high-$z$ clusters resulted in very small improvements,
confirming that diagnostics based on surface brightness gradients 
rather than integral quantities, are more prone to statistical noise.  
Our fitting procedure, based on a Levenberg-Marquardt algorithm, constrain the 
parameter $\beta$ to the range 0.4$<$ $\beta$ $<$1.0.
Our fitting function has the form

\begin{equation}
 S(r) = S_{0} (1+(r/r_{c})^2) ^{-3\beta+0.5} + C
\end{equation}
\label{sbeta}

\noindent where $S_{0}$ is the central surface brightness, $r_{c}$ is
the core radius, and $C$ is a constant background. The correlation
between the best fit $r_{c}$ and $\beta$ values is shown in Figure
\ref{Figbeta} for the 400 SD high-$z$ and RDCS+WARPS samples.  The
apparent correlation between $\beta$ and $r_c$ is due to the
tendency of the $\beta$-model to 
compensate for a larger core radius with a larger beta
However, we notice that the two samples
clearly populate a different $r_{c}$-$\beta$ space. The RDCS+WARPS
clusters prefer smaller core radii and larger $\beta$, in agreement
with our analysis of the $c_{SB}$ parameter distribution.
We checked that the temperature distributions of both samples are 
statistically equivalent, so that the difference among the two samples in the 
$\beta$-$r_c$ plane is not affected by any hidden correlation between the temperature
and the fitting parameters. 

Given the much larger signal-to-noise ratio, local clusters require a
double $\beta$-model fit in order to properly describe the core. 
For the low-z sample our fitting function reads:

\begin{equation}
 S(r) = S_{01} (1+(\frac{r}{r_{c1}})^2) ^{-3\beta_{1}+0.5} + S_{02} (1+(\frac{r}{r_{c2}})^2)^{-3\beta_{2}+0.5} + C
\end{equation}
\label{dbeta}

\begin{figure}[h]
\begin{center}
\includegraphics[width=8.5cm,angle=0]{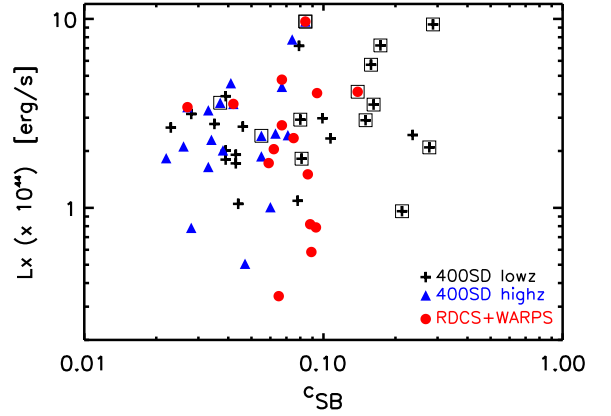}
\end{center}
\caption{X-ray soft-band luminosity $L_{X}$ versus $c_{SB}$. Black crosses refer to the 400 SD 
low-$z$ sample, blue triangles 
represent the 400 SD high-$z$ sample and the red circles correspond to the RDCS+WARPS sample. 
The black squares indicate the presence of a radio source in the cluster center.}
 \label{Figlx}
\end{figure}

The characterization of the surface brightness profile allows us to
measure the total X--ray luminosity of the clusters in our sample.
The contribution of the cool-core to the total cluster luminosity
is known to be, at most, $30$\% in local clusters (see Peres et al. \cite{peres}).
We investigate this effect in our local and distant samples by searching for a 
correlation between the total $L_{X}$ and $c_{SB}$. 
The luminosity excess would be best quantified by a double-$\beta$ model, however, 
given the low significance of such a complex model to describe the low S/N data of 
the distant clusters, we resort to use only total cluster luminosities.
Rest-frame X-ray luminosities in the 0.5-2.0 keV band are computed for
the distant clusters by performing a spectral fit to the inner region
with $r$=40\arcsec (see \S 4), and extrapolating the flux to 1 Mpc
according to the beta-model best-fit parameters.  For the local sample
we used the $L_{X}$ values published in Vikhlinin et al. (\cite{vikhlinin09}).  
In Fig.~\ref{Figlx} we show our results. The three samples span approximately the same range of 
X-ray luminosities (with the distant samples reaching the faintest $L_{X}$), however, 
no clear trend stands out from the $L_{X}$ - $c_{SB}$. relation.
One of the reasons for this result is that, a proper measure of an excess luminosity 
requires a robust characterization of the surface brightness profiles at radii smaller than 
40 kpc, a condition that is hard to fulfill for the high-z cluster sample.

In the following, we will use the results of this Section to derive
the entropy and cooling time of the clusters.

\section{Entropy profiles and central entropy }

Entropy is a fundamental property of the intracluster medium, more so than either 
temperature or density alone. The specific intracluster entropy, $K$, describes the 
thermodynamical history of the gas, and is defined as:

\begin{equation}
K=kTn_{e}^{-2/3}
\end{equation}
\label{dbeta}

\noindent where $T$ and $n_{e}$ correspond to the cluster temperature and gas density 
respectively. 
It is well-known that the entropy profiles of local cool-cores are steeper in the inner 
region, relative to non-cool-cores, reaching values well below 100 keV/cm$^{2}$
(e.g., Donahue et al. \cite{donahue}, Cavagnolo et al. \cite{cav}). 
We measured the cluster entropy profiles using the global cluster temperature to ensure 
a proper comparison between the local and distant samples. 
However, we did investigate the impact of using a single temperature instead of a more 
accurate temperature profile in the core entropy of the local clusters, by exploring 
the database of the ACCEPT\footnote{http://www.pa.msu.edu/astro/MC2/accept/} 
sample (Cavagnolo et al. \cite{cav}). 
The discussion on this comparison is presented in Sect. 7.1. 

\begin{figure}
\begin{center}
\includegraphics[width=8.5cm,angle=0]{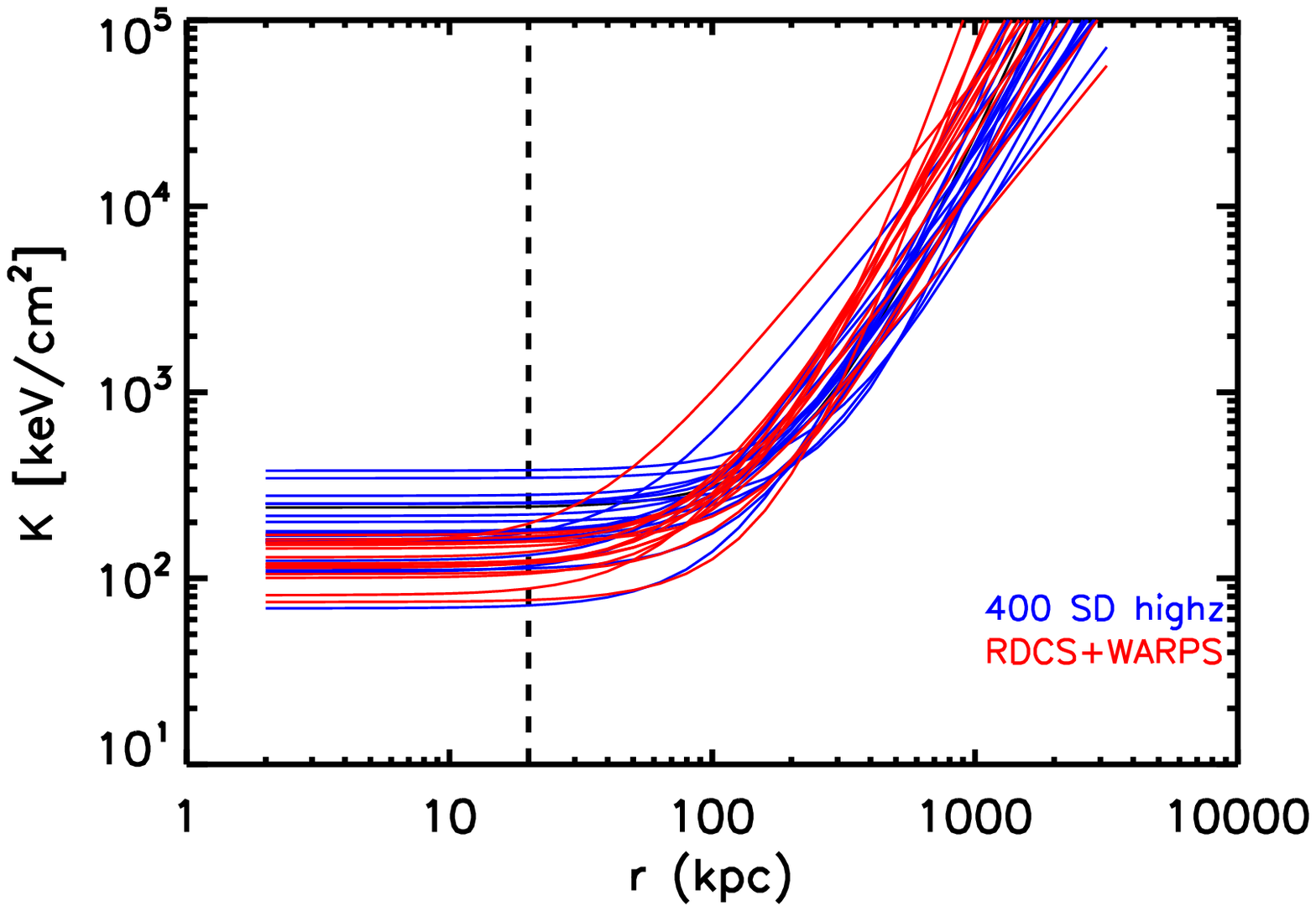}
\includegraphics[width=8.5cm,angle=0]{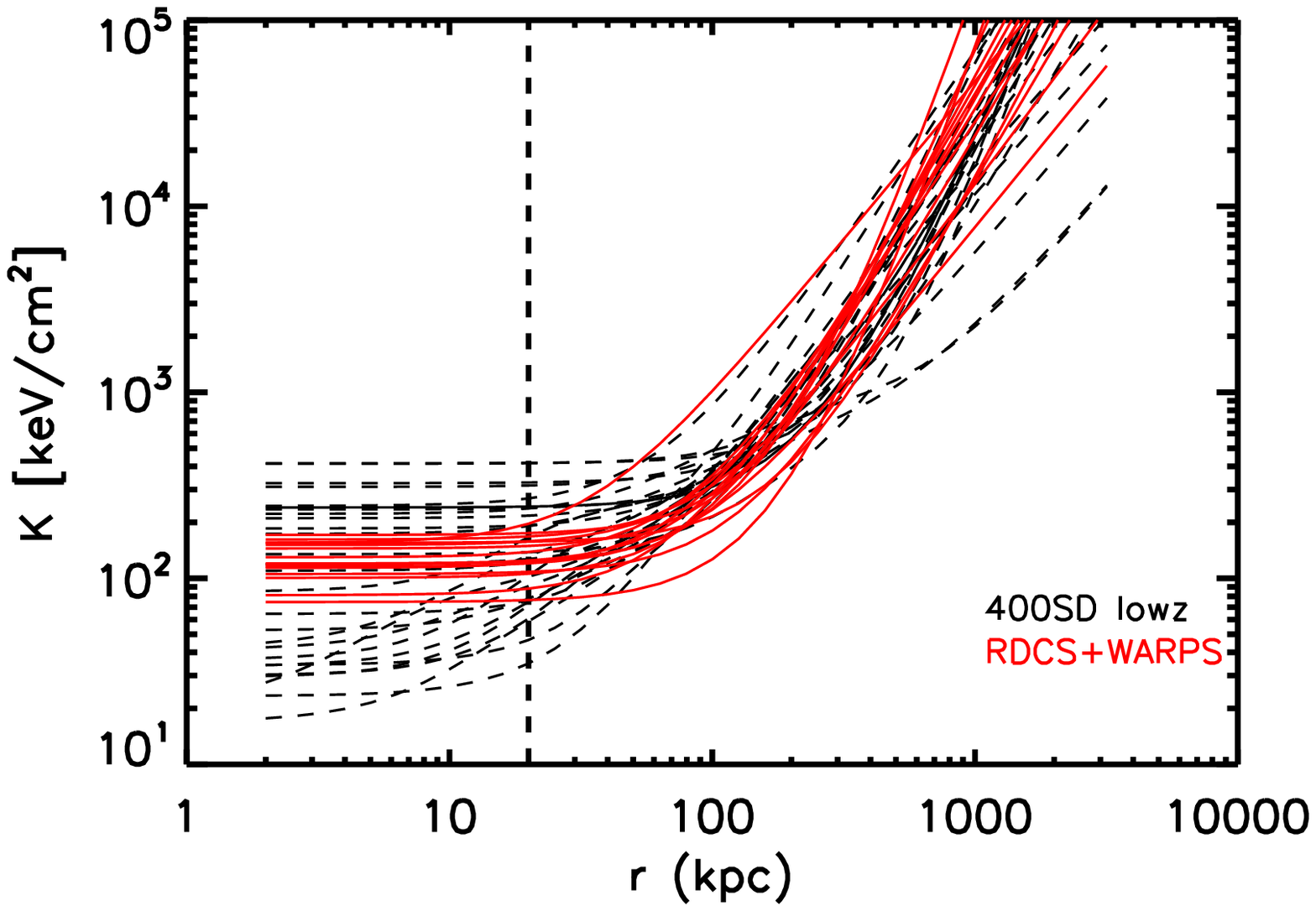}
\end{center}
\caption{Comparison between the entropy profiles of the two distant samples (\textit{top}), 
and the 400 SD low-$z$ relative to the RDCS+WARPS sample (\textit{bottom}). }
\label{kprof}
\end{figure}

The gas density profiles were obtained by deprojecting the surface brightness 
profiles along the line of sight, $S(r) = \int \! n_{e}^{2} \, dl$.
We apply the simple and convenient $\beta$-model approximation to describe the surface 
brightness profiles, and use the best-fit model parameters derived in Section 6.
The entropy profiles obtained in physical units are presented Figure ~\ref{kprof}. 
Again we compare the two distant samples (see top panel of Fig.~\ref{kprof}), but most 
importantly, we compare entropy of the local and the RDCS+WARPS clusters, presented in the bottom 
panel of Fig.~\ref{kprof}.
All profiles are very similar beyond $r$=300 kpc.
The turn over of the profiles takes place at a radius $r$=100-150 kpc, which marks the 
transition from the core region to the cluster outer part.
We have a good agreement with other works that present profiles of distant 
clusters (what is the overlap of clusters), such as Morandi \& Ettori (\cite{morandi}) that present 
entropy profiles of a sample of X-ray luminous clusters with 0.1$<z<$0.8, 
observed with \textit{Chandra}. They find a high similarity between the entropy 
profiles of the low-$z$ and distant clusters, outside the core region. 

The central entropy of the local clusters shows a large spread at very small radii, 
below 20 kpc. This is in part caused by the very steep inner slopes of the profiles 
of cool-core clusters, that can be resolved down to very small scales (2 kpc).

\subsection{Central entropy}

In order to assess and compare the state of the ICM in the core of both the nearby and 
distant clusters, we measured the central entropy at a radius of 20 kpc (K20), a bound 
imposed by the resolution limit of the most distant objects. 
The distribution of K20 for the local and RDCS+WARPS samples is shown in Fig.~10.
To quantify the agreement of these distributions and thus assess the evolution of core 
entropy, we performed a K-S test. 
The probability that the 400 SD low-$z$ and the RDCS+WARPS are associated with the same 
parent distribution is 24\%. In contrast, the 
probability that both 400 SD samples originate from the same distribution is a mere 5\%.
We note however that the local sample spans a much wider range of entropy, relative to the distant clusters: K20 (local)=[35-415] keV/cm$^{2}$, whereas K20 (RDCS+WARPS)=[76-197] keV/cm$^{2}$.

\begin{figure}
\begin{center}
\includegraphics[width=8.5cm,angle=0]{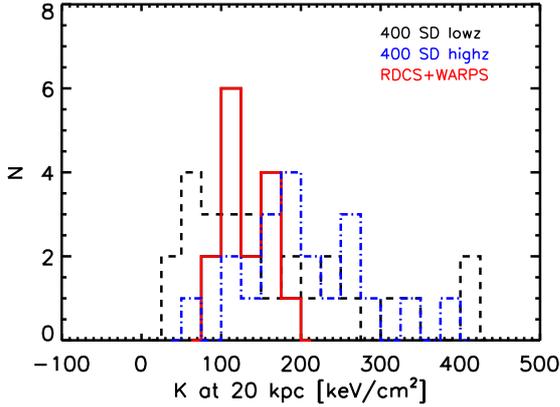}
\caption{Distribution of the entropy measured at 20 kpc. }
\end{center}
\vspace*{-0.3cm}
 \label{k20}
\end{figure}

In addition to comparing the central entropy of the local and distant clusters, 
we also investigated a correlation between K20 and $c_{SB}$.
The result is presented in Fig.~11 and shows a strong anti-correlation between these
two quantities. A Spearman rank test confirms this relation with a coefficient $\rho$=$-$0.84.

\begin{figure}
\begin{center}
\includegraphics[width=8.5cm,angle=0]{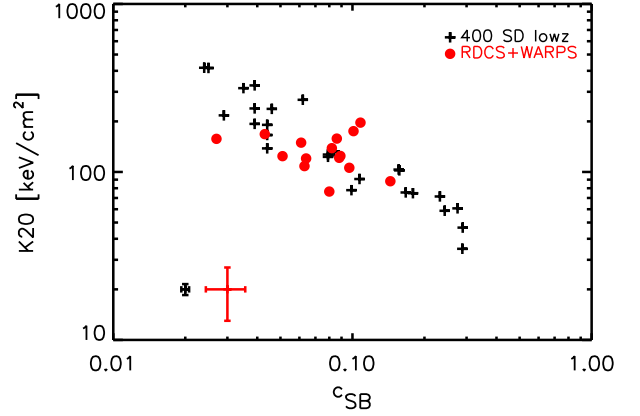}
\caption{Correlation between central entropy K20 and the phenomenological parameter $c_{SB}$. 
 Error bars show typical measurement errors for the low and distant samples.}
\end{center}
\vspace*{-0.3cm}
 \label{kcsb}
\end{figure}

As mentioned earlier, we used the global cluster temperature to compute K20, which 
may introduce a bias in the result, in particular for the local clusters
that are well resolved down to a few kpc. Since all local clusters were independently 
analyzed by Cavagnolo et al., we used the cluster temperatures measured in the innermost bin 
(corresponding to about 10-20 kpc) and compared the histograms of K20 using both the 
central and global temperatures. 
The average difference between the two distributions is 15.3 keV/cm$^{2}$ with a standard 
deviation of 32.0 keV/cm$^{2}$. These distributions differ mainly for the nine clusters with 
K20 $<$100 keV/cm$^{2}$, in which K20 ($T_{global}$) is on average two times higher than K20 ($T_{core}$).
Above K20=100 keV/cm$^{2}$ the difference between K20 ($T_{core}$) and K20 ($T_{global}$) scatters 
around the equality. Therefore, we expect the K20-$c_{SB}$ relation to be even steeper 
for the local clusters, if one uses a resolved central temperature.
We do not find any evidence for bimodality in the distributions of K20.

\section{The central cooling time}

The central cooling time is the measure most often used to quantify cool-cores, 
as it provides a time-frame for the evolutionary state of the gas. 
Adopting an isobaric cooling model for the central gas, $t_{cool}$ can be computed as:

\begin{equation}
t_{cool} = \frac{2.5n_{g}T}{n_{e}^{2} \Lambda(T) }   
\end{equation}
\label{tcool}

\noindent where $\Lambda(T)$, $n_{g}$, $n_{e}$ and T are the cooling function, 
gas number density, electron number density and temperature respectively 
(Peterson \& Fabian \cite{peterson05}), and with $n_{g}$=1.9$n_{e}$.
Using the global cluster temperature and thus considering our results 
as upper limits, we obtained the central cooling time measured
within a 20 kpc radius (see Fig.~\ref{tcool}).
As expected, the $t_{cool}$ distributions of the local and distant samples 
resemble the distributions of K20. The local clusters span the wide range 
of ages [0.7 - 32.6] Gyr, whereas the RDCS+WARPS sample is limited to 
[4.7 - 14.3] Gyr.
For completeness, we report for each sample the fraction of clusters with 
a central cooling time lower than the age of the Universe at the cluster redshift:
in the local sample, 15 out of 26 (58\%); in the RDCS+WARPS, 4 out of 15 (27\%), 
and in the 400 SD high-z, 2 out of 20  (10\%).
However, a more meaningful quantity would be the cooling time normalized to the age of the cluster, 
defined as the time elapsed since the last major merger event. Considering the 
age of the Universe at $z_{obs}$ is misleading, as this is a loose upper bound on the
age of the cluster.

A K-S test to the distributions of $t_{cool}$ in the local and the RDCS+WARPS 
samples indicates a probability of 9\% that these samples derived from the a common 
population. When comparing the local and distant samples from the 400 SD, the probability 
derived with the K-S test decreases to 4\%.

Again we quantified the impact of using a core temperature in the cooling time measurements.
A K-S test to the central cooling time distributions obtained with a core temperature and 
a global temperature indicates that it is likely that both samples originate from 
the same population, with a probability of 67\%.
The central cooling time measured with a resolved temperature is on average 0.3 Gyr 
shorter relative to the measurement with a global temperature. This effect is 
obviously stronger for cool-core clusters. We quantified this effect in clusters 
with central cooling time shorter than 5 Gyr, and we found that in this case, the ratio 
$t_{cool}$($T_{global}$)  / $t_{cool}$($T_{core}$) reaches a factor 1.4.

\begin{figure}[h]
\begin{center}
\includegraphics[width=8.5cm,angle=0]{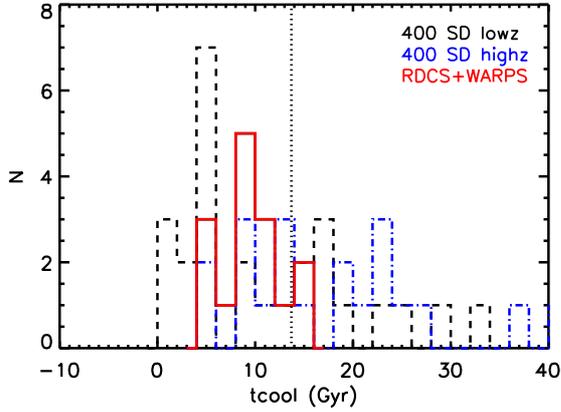}
\end{center}
\caption{Distribution of central cooling time for the local and distant samples. 
The dotted line marks the age of the Universe.}
 \label{tcool}
\end{figure}

\begin{figure}[h]
\begin{center}
\includegraphics[width=8.5cm,angle=0]{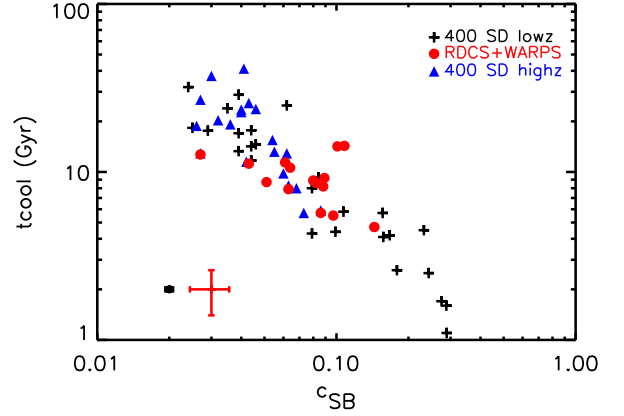}
\end{center}
\caption{Correlation between central cooling time and the phenomenological parameter $c_{SB}$.
Error bars show typical measurement errors for the low and distant samples.}
 \label{tcool2}
\end{figure}

We also find a strong anti-correlation between $t_{cool}$ and 
$c_{SB}$ (see Fig.~\ref{tcool2}), quantified with a Spearman rank test with coefficient
$\rho$=-0.84 and a probability of no  correlation of$<$10$^{-13}$.
This anti-correlation was already shown in Santos et al. (\cite{joana}), and more recently 
Hudson et al. (\cite{hudson}) confirmed it using the nearby HIFLUGCS sample.

\section{Radio sources associated with cluster cores}

In the current scenario, feedback from AGN in the radio mode plays a
prominent role in the physics of cool-cores.  Increasing evidence
from radio and X-ray observations of local clusters show that the
formation of bubbles due to radio jets associated to the central AGN
may effectively satisfy the energy balance between cooling and heating
in cool-core regions (see Blanton \cite{blanton}).  In this respect, the
radio luminosity of the central galaxy can provide an important link
between the black hole activity and the state of the intracluster
medium.

While there are several works pointing out a correlation between the
power associated to the radio activity of the central galaxy to the
cool-core properties (see Sun et al. \cite{sun}),
there are no similar studies at high redshift. 
Using data from the archive of the NRAO VLA Sky Survey (NVSS) at 1.4 GHz 
(Condon et al. \cite{condon}), which is sensitive down to $\sim$ 2.5 mJy, we 
investigated the presence of radio sources in the
cores of our local and distant clusters. The two distant samples are
entirely covered by the NVSS. To determine whether a radio source is
related with the cluster core we adopted a conservative criterion of a
maximum separation of about 20\arcsec between the X-ray centroid and
the center of the radio galaxy. The 400 SD high-$z$ has three radio
sources (3/20), whereas the RDCS+WARPS sample has only 2 out of 15
clusters with a central radio source. This is expected since most of the 
radio sources associated with high-z clusters should be below the flux limit of the NVSS.
Interestingly, the two clusters in the RDCS+WARPS with radio sources
are also considered to be cool-cores with $c_{SB}$ greater than 0.08.

The local sample is not fully covered by the NVSS (coverage of 73\%,
19/26). However, we find a high fraction of clusters with a
radio source (10/19) all associated with high $c_{SB}$ values (i.e.,
greater than 0.08) It is worth noting that most local clusters without
NVSS coverage are non-cool-cores, according to their $c_{SB}$
values. These results are reported in Tables 1, 2 and 3.

\begin{figure}[h]
\begin{center}
\includegraphics[width=8.5cm,angle=0]{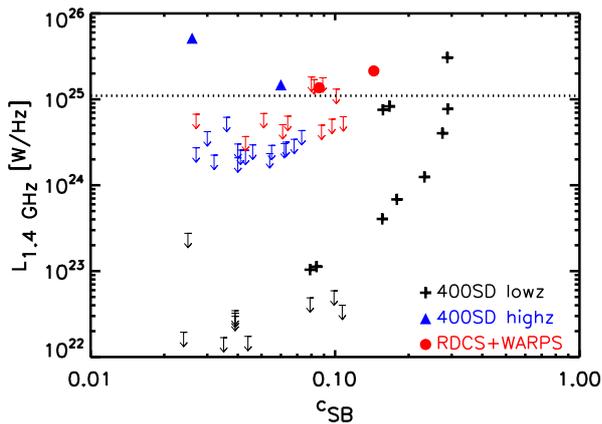}
\end{center}
\caption{Relation between the radio luminosity of sources in close proximity to the 
cluster cores, $L_{1.4 GHz}$, 
and the surface brightness concentration, $c_{SB}$. Arrows refer to upper limits of non-detections.
The dotted line at 1.1$\times 10^{25}$ WHz$^{-1}$ marks the luminosity limit at $z$=1 for NVSS sources, 
corresponding to the flux limit of 2.5 mJy.}
 \label{radio}
\end{figure}

We also obtained the 1.4 GHz luminosity, $L_{1.4 GHz}$, of the radio sources 
associated with the cluster cores, using the fluxes provided by the NVSS catalog, 
and applying a K-correction with a power law with index 0.7. 
We placed upper limits on the non-detections using the NVSS flux detection 
threshold converted into luminosity. 
 As shown in Fig.~\ref{radio}, the $L_{1.4 GHz}$ corresponding to the local 
sample shows a trend with $c_{SB}$, with the stronger cool-cores having 
correspondingly the most luminous radio sources (errors associated with the 
radio luminosities are smaller than 5\%, therefore are not shown). 
To properly take into account the upper limits on the radio luminosity for the 
clusters with NVSS coverage and no detection for the central galaxy we used ASURV 
(Lavalley et al. \cite{lavalley}), the Survival Analysis package which employs the routines 
described in Feigelson \& Nelson (\cite{feigelson}) and Isobe et al. (\cite{isobe}). Survival analysis 
methods evaluate the probability of correlation and linear regression fits by dealing 
properly with non-detections (upper limits). 
We find a very strong (p $>$ 99.0\%) correlation between $c_{SB}$ 
and $L_{1.4 GHz}$ for the local sample, with a Spearman rank test coefficient $\rho$ = 0.82.

The type of radio source usually associated with local cool-cores is the FR-I, with an 
incidence rate of the order of 60\%.
Using the classification made by Fanaroff \& Riley (\cite{fr}) we identified the type 
of radio sources.
 We apply the FR-I/FR-II dividing line provided by Chiaberge et al. (\cite{chiaberge}), 
$L_{1.4 GHz} \sim 4 \times 10^{25}$ W Hz$^{-1}$,
and conclude that, with the exception of the source likely associated with 400d J1221+4918, 
all radio sources belong to the low-luminosity type FR-I. 

The radio sources in the cores of distant clusters populate the high
$L_{1.4 GHz}$ luminosity end, however the small number of detected sources does
not allow us to draw any conclusion. 

The study of radio sources in distant clusters is important to
understand how the feedback mechanism shapes the
evolution of cool-cores, and is particularly useful in searches of
high-$z$ clusters around FR-I sources (Chiaberge et al. \cite{chiaberge}).  However,
the evolution of the radio luminosity function (RLF) remains a
controversial issue requiring further studies. While
Perlman et al. (\cite{perlman04}) finds no evidence for evolution of the radio
luminosity function up to $z$=1, Branchesi et al. (\cite{branchesi}) concludes that the
RLF of distant X-ray clusters is very different from that of local
rich Abell clusters, with a steeper slope at low radio powers
($\le 10^{24}$ WHz$^{-1}$).

\section{Looking into the future: perspectives with WFXT}

With the present work we show that we can explore the population of
cool-core clusters up to the highest redshift where X-ray clusters are
selected (z$\sim1.4$) by fully exploiting the archive of \textit{Chandra}.  
This study is possible thanks to
the exquisite angular resolution of \textit{Chandra}, which allows us to sample
the cool-core region (corresponding to a radius of 40 kpc) at any
redshift with a resolution factor (defined as the physical scale we want to 
resolve divided by the half energy width of the instrument PSF) of about 10.  The only way to
improve the present work is to add serendipitously discovered high
redshift clusters followed-up with deep \textit{Chandra} observations.  
Even though the number of $z>1$ X-ray clusters is slowly increasing thanks
to ongoing surveys as the XMM-LSS; the XMM-Newton Cluster Survey; 
the XMM-Newton Distant Cluster Project; and the Swift X-ray Cluster
Survey,  sample statistics is not expected to increase significantly 
without a dedicated wide area, deep X-ray survey.  Hence, it is instructive to 
look into the future X-ray missions to investigate the capability of characterizing
the cool-core strength of high-$z$ clusters.

Unfortunately, no proposed or planned X-ray facility foresees an
angular resolution comparable to that of \textit{Chandra}. Therefore, any 
future X-ray work on the evolution of cool-cores at high-redshift
must deal with the blurring effect of the PSF on the cluster images.
In order to have low systematics errors, we need to keep the HEW as low as possible.  
As shown in Santos et al. (\cite{joana10}), the capability to
classify strong, moderate and non cool-core clusters up to $z\sim
1.5$ using $c_{SB}$ is strongly dependent on the instrument HEW.
Results obtained with $c_{SB}$ are still reliable when the HEW is 5 arcsec, 
but beyond 10\arcsec it is no longer possible to confidently 
discriminate between a cool-core and a non cool-core cluster.  
Only two proposed X-ray missions expect to have a PSF with a 5 arcsec HEW
at 1 keV: the International X--ray Observatory and the Wide Field
X--ray Telescope.

The International X-ray Observatory (IXO) (see e.g., Bookbinder et al. \cite{bookbinder}) 
is designed to have a great collecting power and high
spectral resolution, therefore it will provide very detailed analysis
of known or serendipitously discovered clusters, up to high-redshift.
However, IXO will not be used in a survey mode for large area surveys, 
and thus it would not significantly increase the statistics of
high-z cluster samples.  

One of the most promising proposed X-ray missions is the Wide Field
X-ray Telescope (WFXT), a medium-class mission designed to be two
orders of magnitude more sensitive than any previous or planned X-ray
mission for large area surveys, and to match in sensitivity the next
generation of wide-area optical, IR and radio surveys.  In five years
of operation, WFXT will carry out three extragalactic surveys: the
deep (400 ksec, 100 deg$^{2}$), medium (13 ksec, 3000 deg$^{2}$) and
wide (2 ksec, 20000 deg$^{2}$) surveys (see Giacconi et al. \cite{giacconi}).  The
combination of good angular resolution across the entire field of view
(HEW $\sim 5$ arcsec), large collecting area and large field of view
(one square degree), makes WFXT the most efficient finding machine of
X-ray clusters at $z>1$.  The expected number of high-z clusters
detected in the WFXT surveys with signal-to-noise comparable to that
of the cluster sample in this work (conservatively expressed as a
lower bound of 1500 net counts), is shown in Table ~\ref{abun}.

\begin{figure*}
\begin{center}
\includegraphics[width=4cm,height=4cm,angle=0]{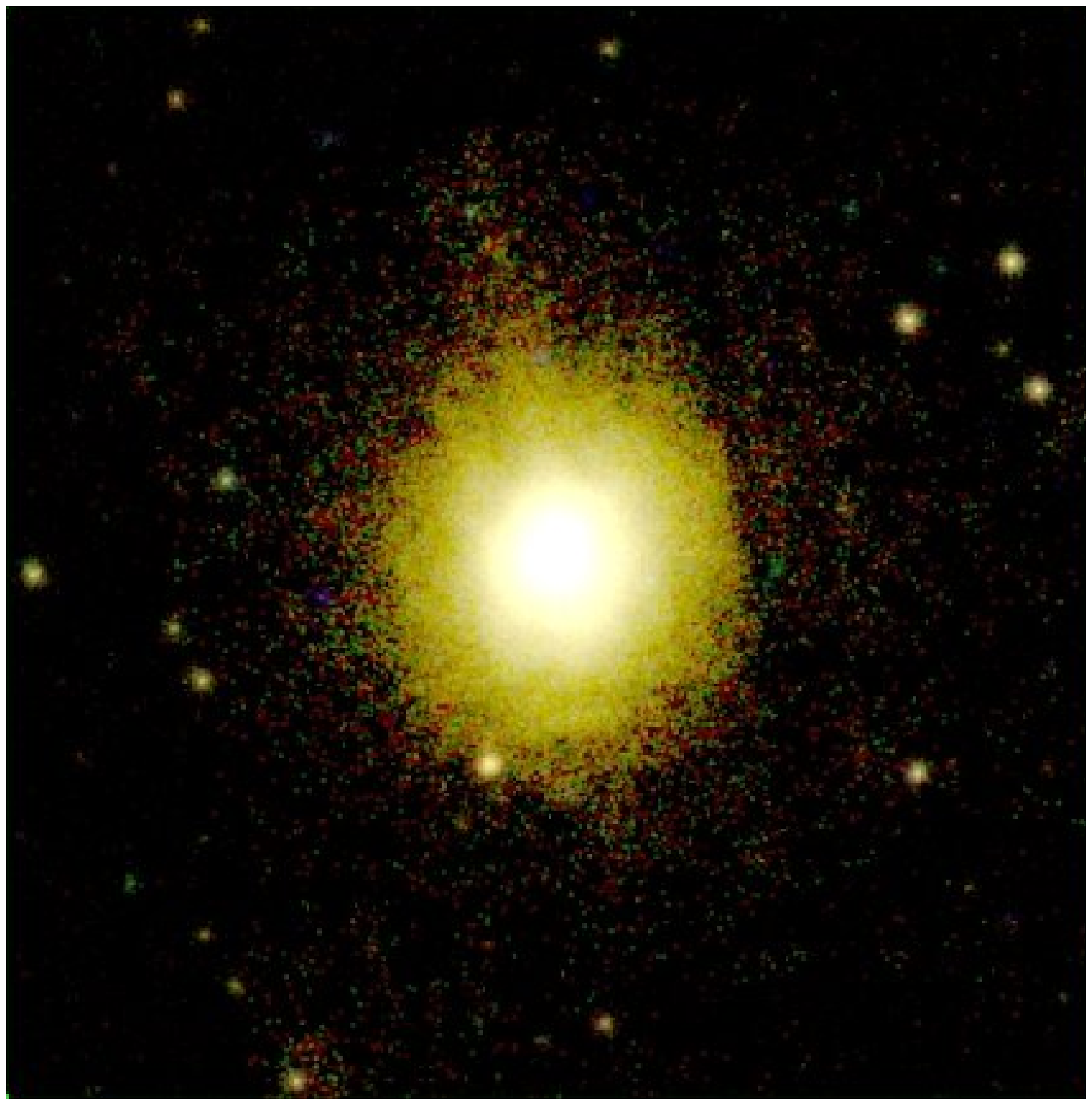}
\includegraphics[width=4cm,height=4cm,angle=0]{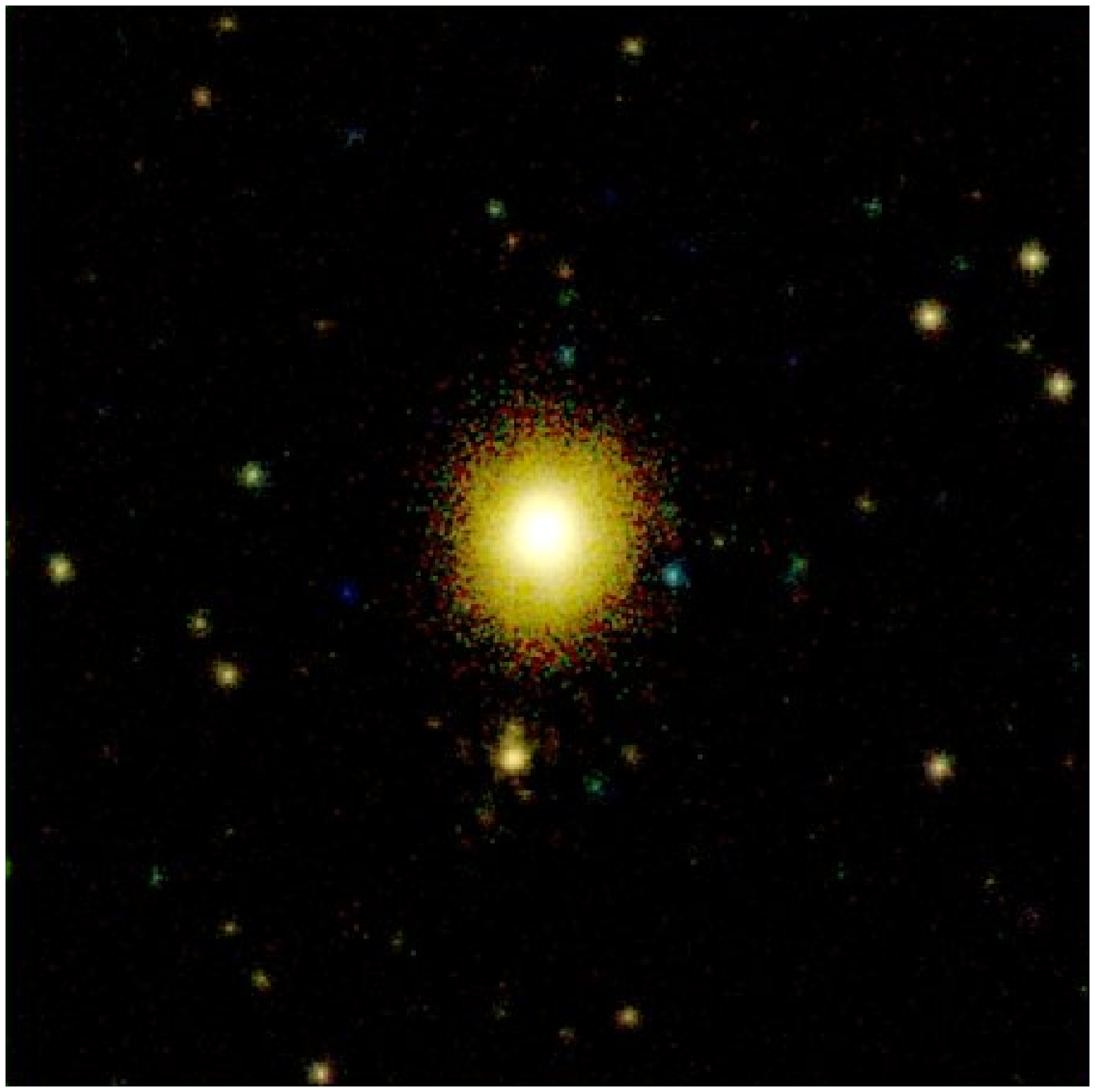}
\includegraphics[width=4cm,height=4cm,angle=0]{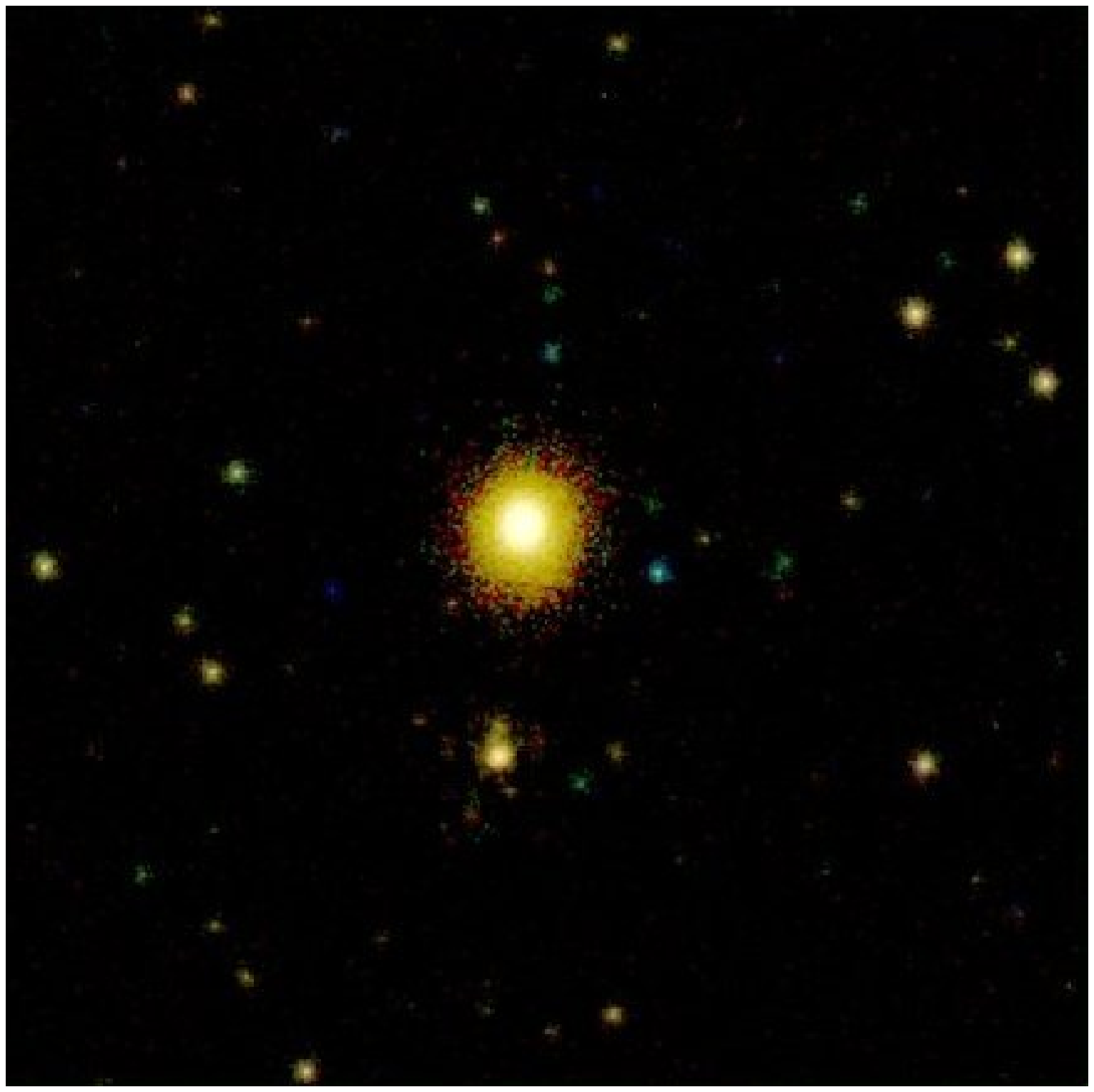}
\end{center}
\caption{Simulated images of strong cool-core A1835 as seen by WFXT
  in medium exposure (13.2 ksec), at redshifts 0.5 (left), 1.0
  (middle) and 1.5 (right).  The images have a size of 10$\times$10
  arcmin and are displayed in logarithmic scale.}
 \label{wfxt}
\end{figure*}

\begin{table}
\caption{Expected number of clusters with temperature $kT >$3 keV and
  with more than 1500 net counts in the 0.5-7 keV band, in each one of
  the three planned WFXT surveys. }
\label{abun}
\begin{center}
\begin{tabular}{lll}
\hline
\\
Survey & 0.5$<z<$1.0 &  1.0$<z<$1.5 \\
\hline
\\
Shallow  & 200   & 0 \\
Medium   & 2190  & 300 \\
Deep     & 188   & 94 \\
\hline
\end{tabular}
\end{center}
\end{table}

The image quality of WFXT allows one to robustly
characterize the cool-core strength in clusters detected with
more than 1500 net counts.  Simulations of realistic WFXT fields
have been produced in order to investigate this science case.
Using the cloning technique (Santos et al. \cite{joana}), we simulated WFXT
images of typical strong, moderate and non cool-core clusters at
redshifts 0.5, 1.0 and 1.5.  The templates chosen to perform the cloning 
simulations are A1835, A963 and A2163, which represent these three cluster types. 

To obtain a quantitative assessment of the cool-core properties of the
simulated clusters, we measured $c_{SB}$.  For the most interesting case of 
tracing the evolution of a strong cool-core (A1835, see Fig.~\ref{wfxt}), 
we measured $c_{SB}= 0.184, 0.156$ and $0.152$ at $z$=0.5, 1.0 and 1.5,
respectively, while the \textit{Chandra} value in the original image is
$c_{SB}=0.246$.  We, therefore, find an apparent evolution due to the
angular resolution, which can be removed by applying a proper
deconvolution of the PSF.  However, and more importantly, we also confirm 
that the measure of $c_{SB}$ at face-value allows us to assign each cluster to its own
cool-core class (strong-, moderate- and non-CC) at any redshift.  We
conclude that, while the exact amount of evolution in the cool-core
strength seen by WFXT is moderately affected by the angular
resolution, we will be able to perform a similar study as presented here with \textit{Chandra}
data.

We note that $z=1.5$ is probably the redshift limit to perform
this kind of analysis.  The power of WFXT can be appreciated if we
look at the number statistics below $z$=1.5 in the medium survey,
where we expect to detect 2190 (300) massive clusters (T$>$3 keV) with
a minimum of 1500 net counts (to allow for a robust $c_{SB}$
measurement and a spectral analysis), in the redshift range 0.5$<z<$1.0 (1.0$<z<$1.5).

\section{Conclusions}

 In this paper we investigated the evolution of cool-core clusters across 
the entire redshift range currently available, i.e., out to $z$=1.3.
Our analysis is mostly based on the archival \textit{Chandra} data of 
three cluster samples, and our results are derived mainly from the cluster 
X-ray surface brightness properties.
In the following, we summarize our main results and conclusions:

\begin{itemize}
 \item the range of surface brightness spanned by the three samples (see
   Figure 1, 2, and 3) includes a wide variety of physical morphologies, 
  illustrating that the samples are well representative of the X-ray 
luminous galaxy cluster population and therefore are suitable to 
investigate the evolution of the cool-cores;
 \item the distributions of the surface brightness concentration,
   $c_{SB}$, in the high-$z$ cluster samples (Figure ~\ref{Figcsb},
   left panel) show that the two samples are statistically different:
   the 400 SD high-$z$ sample lacks concentrated clusters, with median
   $c_{SB}$ equal to 0.043, in contrast, the RDCS+WARPS spans a
   wider range of $c_{SB}$, with median $c_{SB}$=0.082. This finding
   suggests that the 400 SD high-$z$ sample may be biased against
   concentrated morphologies;
\item the comparison between the local sample and the RDCS+WARPS
  clusters shows that the distributions of $c_{SB}$ are not
  significantly different (see Figure ~\ref{Figcsb}, right panel),
   exposing a well defined population of cool-cores already in
  place at $z\sim 1.3$; however, the high-$z$ sample is deficient in very
  peaked (or strong cool-core, with $c_{SB}$ $>$0.15) clusters;
 \item the entropy profiles of the low and high-$z$ samples agree well
   in the outskirts but show a large scatter in the inner regions; the
   comparison between the distributions of the entropy value at 20 kpc (K20)
 in the local and the
   RDCS+WARPS samples shows that, even though both distributions share
   the same median value (127 and 125 keV/cm$^{2}$ in the local and
   distant samples, respectively) the distant sample covers a narrower
   range of central entropy, K20=[76-197]keV/cm$^{2}$, and does not
   reach values as low as the local sample.  
 \item we measured the cluster central cooling times at 20 kpc radius 
  and found that the local sample spans a broad range of cooling time
   ([0.7-32.6] Gyr), with a median value of 9.3 Gyr, and a significant peak at $t_{cool}$=5 Gyr.  
   The RDCS+WARPS sample shows a somewhat different
   behaviour, displaying a narrower range of cooling times,
   [4.7-14.3] Gyr, and a median $t_{cool}\sim$ of 8.9 Gyr. 
   Finally, we confirm the strong anti-correlation
   between $c_{SB}$ and $t_{cool}$, mirrored by the anticorrelation between 
   $c_{SB}$ and K20 (see Figure ~11); 
 \item our results on $c_{SB}$, central entropy and central cooling time do not support 
recent claims of a cool-core/non cool-core bimodality;
 \item using the NVSS archive we identified radio sources in close vicinity with the 
cluster cores ($<$ 20\arcsec).
We find a significant correlation between $c_{SB}$ and the 
radio luminosity of sources in the local sample. All local 
clusters with a central radio source are cool-cores, since they have $c_{SB}\ge$0.80.
The lack of radio data for the distant clusters does not allow us to draw any 
conclusion on the connection between the cool-core strength and the presence of 
a central radio galaxy (Fig.~\ref{radio}), which motivates us to explore this line of research 
in the future, using deep, high-resolution radio data, to investigate the interplay between 
the X-ray core, the brightest cluster galaxy and the central radio source, in high-redshift clusters. 
\end{itemize}

Our findings can be used to significantly constrain the formation time-scale 
 of cool-cores. We note that one of the most distant clusters discovered to date 
and not included in our high-z sample, XMMUJ2235 at $z$=1.393 (Mullis et al. \cite{mullis}) 
found within the framework of the XDCP (B\"ohringer et al. \cite{boehringer05}), was also 
analysed in Santos et al. (\cite{joana}) and Rosati et al. (\cite{rosati09}). We measured 
$c_{SB}$ equal to 0.103, confirming XMMUJ2235 as a cool-core cluster.

The results presented in this paper extend the current knowledge of the cool-core 
population to the most distant clusters known, and shows us that even at such large lookback 
times, we still find a large population of moderate cool-core clusters. A direct consequence 
is that a significant number of cool-cores develops on times scale comparable
to the dynamical time scale of clusters ($\sim 1$ Gyr).

With this work we are reaching the limit of what can be done with current X-ray 
missions to study the cores of the most distant galaxy clusters. The small number 
statistics of high-$z$ clusters is the major limitation that prevents us from a more 
accurate understanding of the physics of the youngest clusters, although an equally 
important requirement for such a study is a high angular resolution, as the one reached 
by \textit{Chandra}. 
A complementary strategy is to go as deep as possible with \textit{Chandra} on high-$z$ 
cool core clusters.  In this way, we will exploit at best the presently limited sample of 
high-$z$ clusters, in order to unveil the properties of their cool cores (temperature and 
metallicity profiles, presence of cavities associated to radio sources) and compare them to 
those in local clusters.  This can be achieved only with very deep (several hundreds of ksec) 
Chandra pointings.

 A significant advancement in this research will be achieved when large samples are available,
 which will only be possible with the next generation X-ray survey missions.
In particular, we showed that the next generation X-ray survey instruments with a 
large collecting area and good angular resolution will have the ability to 
resolve the central regions of strong cool-cores up to a redshift of 1.5.

\begin{acknowledgements}
      We wish to thank S. Molendi, S. Borgani and S. Ettori for useful discussions in 
      various stages of this study.
      We acknowledge support from ASI-INAF I/088/06/0. PT aknowledges support
under the grant INFN PD51. PR acknowledges partial support by the DFG cluster of 
excellence Origin and Structure of the Universe (http://www.universe- cluster.de).
\end{acknowledgements}

\bibliographystyle{aa}

\end{document}